\documentclass[onecolumn,authoryear]{els-mrw} 

\usepackage{amsmath,amssymb,amsfonts,amsthm,makeidx,graphicx}
\usepackage{txfonts}
\usepackage{helvet}
\usepackage[greek, english]{babel}

\usepackage{textcomp}
\providecommand\micron{\mbox{\textmu{}m}}%

\begin{document}

\chapter{Minor planets, asteroids, comets and interplanetary dust within 30 au}\label{chap1}

\author[1,2]{Quanzhi Ye}%

\address[1]{\orgname{University of Maryland}, \orgdiv{Department of Astronomy}, \orgaddress{4296 Stadium Dr, College Park, MD 20742, USA}}
\address[2]{\orgname{Boston University}, \orgdiv{Center for Space Physics}, \orgaddress{725 Commonwealth Ave, Boston, MA 02215, USA}}

\articletag{Chapter Article tagline: update of previous edition,, reprint..}

\maketitle

\begin{glossary}[Learning objectives]

\begin{itemize}
    \item Understand the different types of small bodies and their interrelations;
    \item Learn about the asteroid--comet continuum;
    \item Learn about the orbital and compositional structure, formation, and evolution of the main asteroid belt and the Jupiter Trojans;
    \item Understand the origin of near-Earth objects as escapees from the asteroid belt, and recognize their impact hazard to Earth;
    \item Learn about Centaurs as an intermediate stage between the trans-Neptunian objects and Jupiter-family comets;
    \item Know the orbital and physical characteristics of different comet populations, and understand their connection to the interplanetary dust;
    \item Understand the structure and evolution of the interplanetary dust cloud;
    \item Recognize the importance of small bodies in exploring planetary formation in the Solar System and planetary systems in general.
\end{itemize}

\end{glossary}

\begin{glossary}[Glossary]
\term{Active asteroid.} -- An object on an asteroid-like orbit that exhibits comet-like activity.\\
\term{Asteroid.} -- A small rocky body that orbit the Sun.\\
\term{Bolide.} -- An exceptionally bright meteor with an apparent magnitude of $<-14$.\\
\term{Centaur.} -- A small body that orbit the Sun between the orbits of Jupiter and Neptune\\
\term{Chondrite.} -- A type of primitive meteorite that has not been significantly altered (by melting or differentiation) since its formation.\\
\term{Comet.} -- A small ice-rich body in the Solar System that can exhibit a coma and/or a tail.\\
\term{Fireball.} -- A very bright meteor with an apparent magnitude of $<-4$.\\
\term{Interplanetary dust.} -- Dust particles within planetary systems.\\
\term{Micrometeorite.} -- A small ($\lesssim 10~\micron$) interplanetary dust particle that does not produce meteors and can survive entry through Earth's atmosphere.\\
\term{Meteor.} -- The luminous phenomena resulting from meteoroids entering a planetary atmosphere.\\
\term{Meteorite.} -- Macroscopic remnant of an asteroid or a large meteoroid that reaches the surface of a planet or moon.\\
\term{Meteoroid.} -- Larger interplanetary dust that produce meteors and impact flashes.\\
\term{Minor planet.} -- An object that orbits the Sun that is neither a planet nor a comet.\\
\term{Planetary defense.} -- The effort of investigating and mitigating the risks posed by near-Earth objects.\\
\term{Planetesimals.} -- Building blocks of planets formed in the early stage of planetary formation.\\
\term{Resonance.} -- Situation where gravitational influence from a nearby planet is enhanced, leading to highly stable or unstable interaction.\\
\term{Small (Solar System) body.} -- A natural solar System object that is neither planet or dwarf planet, nor a moon. \\
\term{Volatiles.} -- Chemical elements or compounds with low vaporization points.\\
\term{Yarkovsky effect.} -- Modification of the orbit of a small body caused by anisotropic thermal emission.\\
\term{YORP effect.} -- Modification of the spin state of a small body caused by anisotropic thermal emission.
\end{glossary}

\begin{glossary}[Nomenclature]
\begin{tabular}{@{}lp{34pc}@{}}
ACO &Asteroid on a cometary orbit\\
au &Astronomical Unit (the mean Sun-Earth distance, or $1.5\times10^{11}$~m)\\
ETC &Encke-type comet\\
HTC &Halley-type comet\\
IAU &International Astronomical Union\\
IDP &Interplanetary dust particle\\
JFC &Jupiter-family comet\\
LPC &Long-period comet\\
MPC &Minor Planet Center\\
$M_\odot$ &Mass of the Sun, or $2.0\times10^{30}$~kg\\
MOID &Minimum orbit intersection distance\\
NEA &Near-Earth asteroid\\
NEO &Near-Earth object\\
PHA &Potentially hazardous asteroid\\
PHO &Potentially hazardous object\\
$q$ &Perihelion distance\\
$Q$ &Aphelion distance\\
SPC &Short-period comet\\
TNO &Trans-Neptunian object\\
YORP &Yarkovsky–O’Keefe–Radzievskii–Paddack (effect)\\
\end{tabular}
\end{glossary}

\begin{abstract}[Abstract]
Our Solar System includes the Sun, eight major planets and their moons, along with numerous asteroids, comets, and dust particles, collectively known as the small Solar System bodies. Small bodies are relics from the birth of the Solar System and offer valuable insights into planetary formation and the origins of life. This chapter explores this important component of our Solar System, discussing the formation and evolution of key small body populations and their interrelations.
\end{abstract}

\section{Introduction}

When we look up at the night sky, our attention is immediately drawn to the most obvious celestial objects: the Moon, the planets, bright stars, and (under a sufficiently dark sky) the Milky Way. If we watch long enough and carefully enough, we may notice other spectacles: meteors (shooting stars) streaking across the sky, the faint band of zodiacal light roughly marking the ecliptic, and, once every few years, a bright comet lingering for a while.

These objects and phenomena show that our Solar System contains much more than just the Sun, the Moon, and the planets. In fact, small Solar System bodies -- loosely defined as minor planets, asteroids, comets, and interplanetary dust -- are the most numerous objects in the Solar System. As of mid-2024, approximately 1.4 million minor planets (including asteroids) and 5,000 comets have been cataloged. The total number of small bodies of 1~km or larger is in the billions.

\begin{figure}
\centering
\includegraphics[width=\textwidth]{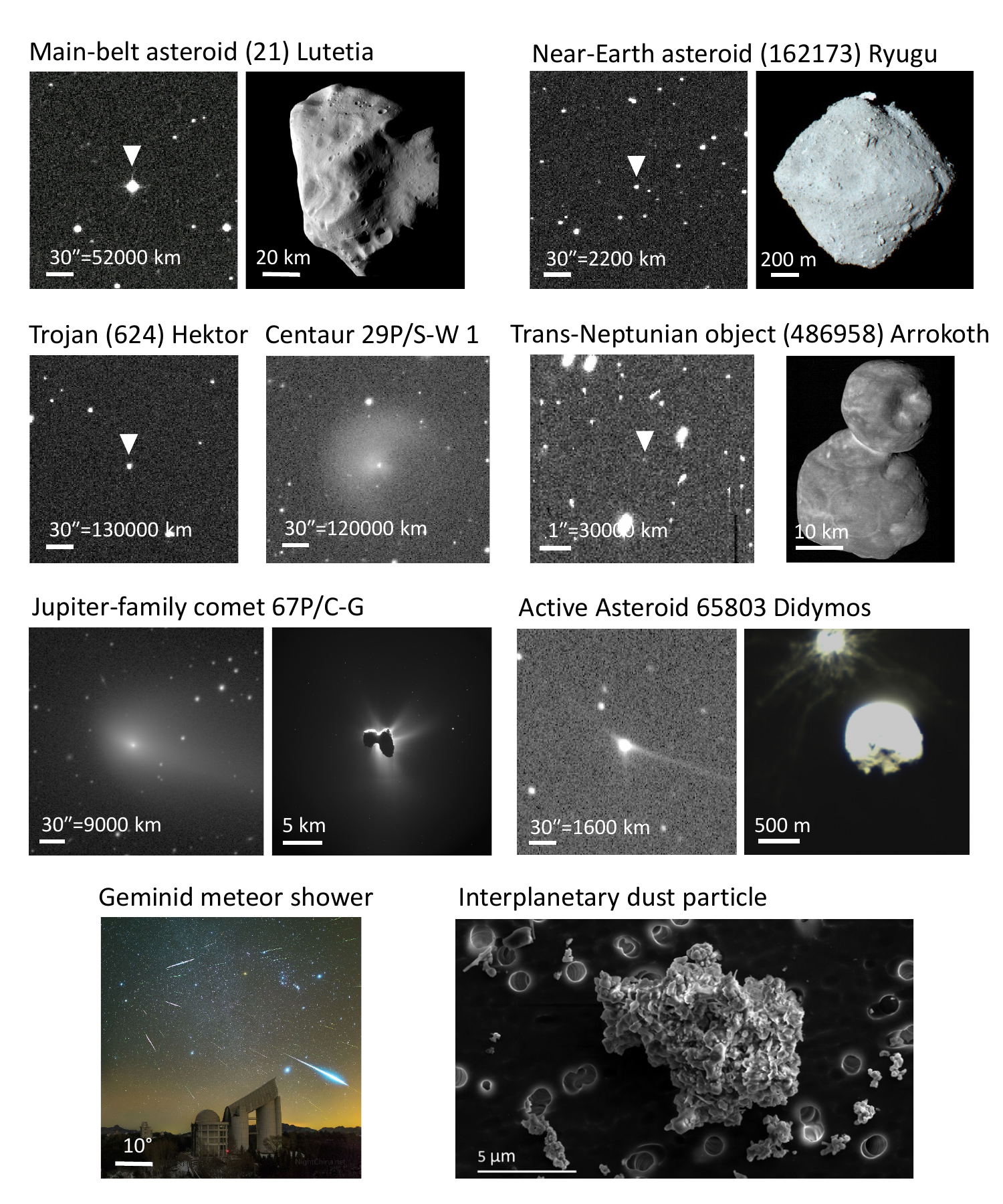}
\caption{A medley of objects discussed in this chapter: main-belt asteroid (21) Lutetia (left: ground-based image taken by the Palomar Transient Factory; right: in-situ image taken by the Rosetta spacecraft, ESA/MPS/UPD/LAM/IAA/RSSD/INTA/UPM/DASP/IDA/OSIRIS team); near-Earth asteroid (162173) Ryugu (left: image taken by the Zwicky Transient Facility/ZTF, right: in-situ image taken by the Hayabusa-2 spacecraft, ISAS/JAXA); Jupiter Trojan (624) Hektor; Centaur 29P/Schwassmann--Wachmann 1 during an outburst; Trans-Neptunian object (486958) Arrokoth (left: image taken by the Hubble Space Telescope, NASA, ESA, SwRI, JHU/APL, the New Horizons KBO Search Team; right: in-situ image taken by the New Horizons spacecraft, NASA/APL/SwRI/NOAO); Jupiter-family comet 67P/Churyumov-Gerasimenko (left: image taken by ZTF; right: image taken by the Rosetta spacecraft, ESA/Rosetta/NAVCAM); near-Earth asteroid (65803) Didymos during and after the impact of the Double Asteroid Redirection Test/DART spacecraft (left: ground-based image taken by ZTF 4 days after the impact; right: in-situ image taken by LICIACube satellite during the impact, NASA/ASI); the Geminid meteor shower (Steed Yu); and an interplanetary dust particle (Hope Ishii, University of Hawai`i at M\=anoa). Scales shown for the Earth-based images are the projected lengths of the scale bars at the distance of the objects.}
\label{fig:objects}
\end{figure}

\begin{figure}[t]
\centering
\includegraphics[width=\textwidth]{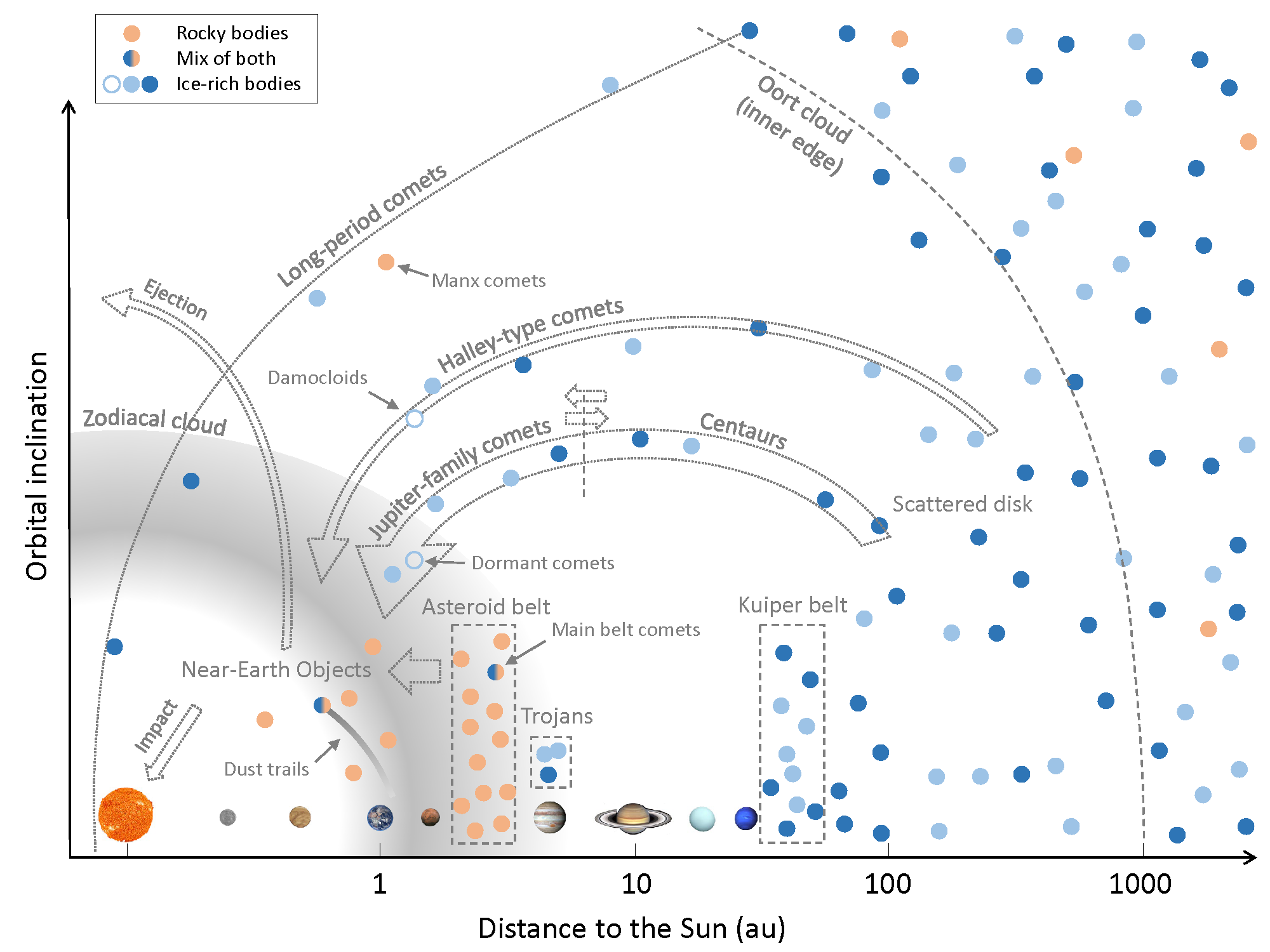}
\caption{Key small body populations and their interrelations.}
\label{fig:schematic}
\end{figure}

Great numbers come with great diversity. The classical terms ``asteroids'' (rocky bodies) and ``comets'' (ice-rich bodies) are now recognized as observable endpoints on a continuum of objects with a wide range of observational, compositional, and dynamical properties. These objects can shift along this continuum as they evolve and interact with other bodies. Studying them provides valuable insights into planetary formation and evolution, the potential threats posed by near-Earth asteroids and comets, and the origins of life and life-essential materials such as water. As a teaser, Figures~\ref{fig:objects} and \ref{fig:schematic} show different types of small bodies and a schematic diagram of their interrelationships.

In this chapter, we will explore the various small body populations within the Solar System, from the Sun outward to the trans-Neptunian region. This boundary is chosen partly because most structures and processes we can currently explore lie within it. We begin by defining different types of small bodies -- minor planets, asteroids, comets, interplanetary dusts -- and discussing key concepts as well as the asteroid--comet continuum (\S~\ref{sec:types}). We will then delve into each small body population: the asteroid belt, Jupiter Trojans, near-Earth objects, Centaurs, the various comet populations, and the interplanetary dust cloud (\S~\ref{sec:population}). Finally, the chapter concludes with a summary of recent advances and future prospects of small body science (\S~\ref{sec:conclusion}).

\section{Types of Objects}
\label{sec:types}

\subsection{Definitions and Key Concepts}

The population of small bodies encompasses a wide range of object types, which are sometimes loosely defined. Small body science also involves concepts that are not commonly used in other areas of astronomy. Therefore, we will begin by reviewing the different types of small bodies and discussing several key concepts.

\paragraph{Minor Planets}

The term \textit{minor planets} is an umbrella term covering all natural macroscopic objects that orbits the Sun but are neither planets nor comets. This term was coined in the 1840s and was initially used interchangeably with the term ``asteroids'', as minor planets known at that time were all located in the asteroid belt. The two terms started to diverge in the 1920s with the discovery of minor planets beyond Jupiter's orbit. Since then, the term ``minor planet'' generally encompasses asteroids within Jupiter's orbit as well as nearly all objects beyond, except for comets.

In 2006, the International Astronomical Union (IAU) introduced a new definition of planets, partly to address whether Pluto should be considered as a planet. Under this new definition, Pluto was reclassified as a \textit{dwarf planet}, a new category that falls under the broader definition of minor planets. Pluto was assigned a permanent minor planet number, 134340, along with three other dwarf planets already recognized as minor planets: (1) Ceres, (90377) Sedna, and (136199) Eris.

\paragraph{Small Bodies}

While minor planets exclude comets by definition, the line between the two can be nebulous (see \S~\ref{sec:types-continuum}). Since the 1970s, the term \textit{small (Solar System) bodies} has been commonly used to collectively describe both populations. The 2006 IAU resolution defined a small body as an object in the Solar System that is neither a planet, dwarf planet, nor a moon, effectively including most minor planets and comets. Usage varies on whether small bodies include interplanetary dust.

Despite the term ``minor planet'' in its name, the IAU Minor Planet Center (MPC) is the official organization responsible for collecting observational data on all types of macroscopic small bodies and natural moons, not just minor planets. The MPC also calculates the orbits of these objects and assigns official designations.

\paragraph{Asteroids}

The term \textit{asteroids} originally referred to bodies found in the asteroid belt but has since been expanded to include any non-cometary bodies that orbit the Sun, particularly those within the orbit of Jupiter. The word is derived from the Greek word \textgreek{ἀστεροειδής}, meaning ``star-like'', which accurately described their appearance in telescopes.

\paragraph{Comets}

A \textit{comet} is an object that exhibit a coma (an extended atmosphere surrounding the nucleus) and/or tail(s). The word originates from the Greek word \textgreek{κομήτης}, meaning ``wearing long hair'' and vividly describes the appearance of these objects.

\paragraph{Interplanetary Dust, Meteoroids, and Meteorites}

\begin{figure}[t]
\centering
\includegraphics[width=\textwidth]{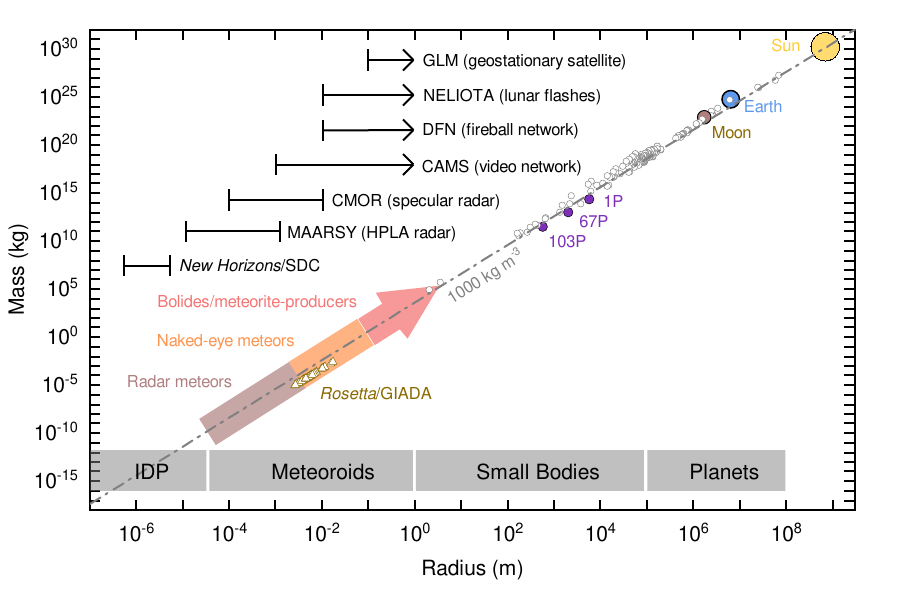}
\caption{A diagram of mass and sizes of various Solar System bodies with a focus of dust/meteor-related phenomena. Also shown are the size regimes of radar, naked-eye (photographic/video) techniques, bolides/meteorite-producers, and major meteor observation programs: the Geostationary Lighting Mapper (GLM), Near-Earth object Lunar Impacts and Optical TrAnsients (NELIOTA), the Desert Fireball Network (DFN), the Cameras for All-Sky Meteor Surveillance (CAMS), the Canadian Meteor Orbital Radar (CMOR), the Middle Atmosphere Alomar Radar System (MAARSY) for High Power Large Aperture (HPLA) radar, and the Student Dust Counter (SDC) aboard New Horizons for space-based dust detectors. Reproduced from \citet{ye2022comets} in \textit{Comets III} \copyright 2024 The Arizona Board of Regents. Reprinted by permission of the University of Arizona Press.}
\label{fig:meteor-def}
\end{figure}

Objects smaller than $\sim1$~m encompass various dust and meteor-related populations (Figure~\ref{fig:meteor-def}), such as:

\begin{description}
    \item[Interplanetary dust.] This term refers to all dust particles within the Solar System. While it is sometimes used interchangeably with \textit{interplanetary dust particles} (IDPs), the latter term often specifically refers to cosmic-origin dust particles collected on Earth. Interplanetary dust mostly originate from macroscopic small bodies such as active and defunct comets, as well as asteroids. These particles produce a wide range of phenomena, including the zodiacal light, meteors, and impact flashes. Interplanetary dust can be studied through various methods, including sample collection from Earth's surface and atmosphere, space missions, interactions between the dust particles and planetary atmospheres or surfaces, and remote sensing in optical and infrared wavelengths. Interplanetary dust is naturally connected to interstellar dust (particles of interstellar origin) which are collectively called \textit{cosmic dust}.
    \item[Zodiacal light and gegenschein.] Zodiacal light is the projection of the zodiacal cloud onto the celestial sphere. Visually, it appears as a faint band of diffuse light along the ecliptic, brighter in directions closer to the Sun and at the antisolar point (opposition) known as the \textit{gegenschein} (German of ``counter-shine''). Gegenschein is caused by the backscattering enhancement of the interplanetary dust cloud by the sunlight. 
    \item[Meteoroids, meteors and impact flashes.] Meteoroids are larger interplanetary dust with sizes ranging from $\sim30~\micron$ to about a meter. Meteoroids at the small end are sometimes called micrometeoroids. The boundary between large meteoroids and small asteroids is nebulous, with a cutoff at 1~m often being used. When meteoroids collide with planets or moons, they can produce \textit{meteors} (the luminous phenomena resulting from meteoroids entering Earth's atmosphere) and \textit{impact flashes} (the luminous phenomena resulting from meteoroids colliding with airless planetary bodies). Meteors with apparent magnitudes of $<-4$ and $<-14$ are called \textit{fireballs} and \textit{bolides}, respectively. Bright bolides may result in meteorite falls. Meteoroids from the same comet or asteroid share similar orbits, forming a \textit{meteoroid stream} that can produce a \textit{meteor shower} when Earth passes through it.
    \item[Meteorites.] A meteorite is a macroscopic remnant of a rocky object from outer space that has landed on the surface of a planet or moon. Meteorites on the Earth usually result from the impact of asteroids of meter-sized or larger, as smaller objects tend to burn up completely in the atmosphere. Meteorites can be classified in several ways. A traditional scheme categorizes them into stony (94\% of all known meteorites), irons (5\%), and stony-iron (1\%) meteorites. \textit{Chondrites} in the stony meteorites is the dominant class of meteorites (86\% of all known meteorites), representing primitive, unaltered material formed during the birth of the Solar System. While most known meteorites are found on Earth, a small number of meteorites have been found on the surfaces of Moon and Mars.
    \item[Micrometeorites.] A micrometeorite is a small IDP ($\lesssim 10~\micron$) that survives entry through Earth's atmosphere. Despite the name, micrometeorites differ significantly from meteorites in terms of composition, origin, and the processes they undergo. Due to their small sizes, these particles are effectively decelerated before they reach meteor altitude ($80-120$~km above the sea level), preventing them from producing visible meteors. Instead, they fall slowly to the surface without undergoing significant changes. Most micrometeorites are compositionally compatible with a sub-class of chondrites known as \textit{carbonaceous chondrites} (4\% of all known meteorites) which are thought to originate from comets.
\end{description}

\paragraph{Dust Dynamics and Evolution}

Compared to macroscopic objects such as asteroids and comets, whose dynamics are largely controlled by the gravity of the Sun and planets, the dynamics of dust are influenced by both gravity and solar radiation. The ratio of the solar radiation pressure to solar gravity, termed $\beta$, is an important measure of dust properties:

\begin{equation}
    \beta = \frac{F_\mathrm{rad}}{F_\mathrm{grav}} = 5.7\times10^{-4} \frac{Q_\mathrm{pr}}{\rho r}
\end{equation}

\noindent where $Q_\mathrm{pr}$ is the efficiency factor of the radiation pressure acting on the particle averaged over the solar spectrum, $\rho$ is the bulk density of the particle, and $r$ is the radius of the particle, all in SI units. Particles larger than the effective wavelength of sunlight have $Q_\mathrm{pr}\sim1$, while smaller particles have $Q_\mathrm{pr}<1$, though the exact value depends strongly on the particle's properties. A particle with $\beta=1$ (approximating a $\micron$-sized grain) and zero initial speed with respect to its parent body experiences equal gravitational and radiation forces in opposite directions, causing it to travel straight out of the Solar System.

\begin{figure}[t]
\centering
\includegraphics[width=\textwidth]{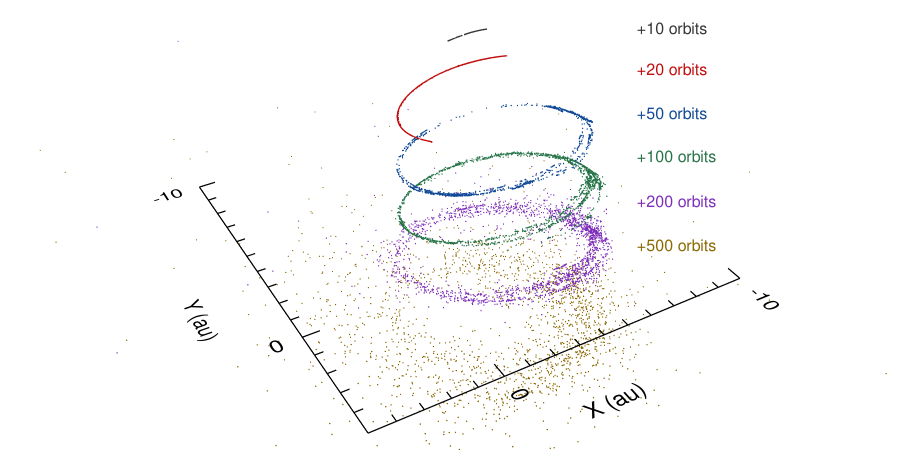}
\caption{Dynamical evolution of mm-sized dust ejected by a Jupiter-family comet. It takes a few $10^2$ orbits for a particle to dynamically blend into the interplanetary dust cloud. Reproduced from \citet{ye2022comets} in \textit{Comets III} \copyright 2024 The Arizona Board of Regents. Reprinted by permission of the University of Arizona Press.}
\label{fig:dust-ev}
\end{figure}

Once released, a particle's dynamical path is modified by gravitational perturbation from major planets and solar radiation effects. It takes a few $10^2$ orbits for a particle to lose the dynamical signature of its parent body and blend into the interplanetary dust background -- the zodiacal cloud (\S~\ref{sec:zodi}). For dust particles released by short-period comets, the time scale of this process is a few kyr (Figure~\ref{fig:dust-ev}).

Dust particles break up over time due to collisions with other dust particles, thermal fatigue, and bombardments of high-energy particles. These processes produce a zodiacal cloud dominated by small, sub-mm-class particles, although the specifics of this process are still not well understood. Meteor observations suggest that the collisional lifetime of mm-class meteoroids at 1~au is on the order of $10^5$ to $10^6$~yr \citep{jenniskens2016cams}.

\begin{BoxTypeA}[box:name]{}

\section*{Poynting-Robertson drag}

The \textit{Poynting-Robertson} drag describes the anisotropic re-emission of photons from a moving particle, which reduces the particle's angular momentum and causes it to spiral slowly towards the Sun. This drag is most pronounced on smaller grains. For example, particles with $\beta=0.1$ takes about 10~kyr to spiral into the Sun from a circular orbit with $a=1$~au.

\end{BoxTypeA}

\paragraph{Absolute Magnitude}

The absolute magnitude is perhaps the most used physical quantity for small bodies. It is defined differently than in other areas of astronomy. For asteroids, absolute magnitude $H$ is defined as the apparent magnitude of the object under a hypothetical condition: the object is at 1~au from both the Sun and the observer, with a phase angle of $\alpha=0^\circ$. Mathematically, 

\begin{equation}
\label{eq:h}
    H = m - 5\log{(r_\mathrm{H} \varDelta)} - \Phi(\alpha)
\end{equation}

\noindent where $m$ is the apparent magnitude, $r_\mathrm{H}$ and $\varDelta$ are the distances to the Sun and the observer in au, and $\Phi(\alpha)$ is the \textit{phase function} that describes the phase-dependent reflectance of the object and is related to surface properties.

For comets, two magnitudes are used: the \textit{total magnitude} $m_1$, which refers to the brightness of the entire coma, and the \textit{nuclear magnitude} $m_2$, which refers to the brightness of the visible core region (not necessarily the comet's nucleus). Total magnitudes are more commonly used, as the definition of ``nucleus'' in the $m_2$ system can differ between observers and instruments. The absolute total magnitude, $M_1$, is defined as

\begin{equation}
    M_1 = m_1 - 2.5 n \log{r_\mathrm{H}} - 5\log{\varDelta} - \Phi(\alpha)
\end{equation}

\noindent where $n$ is sometimes called the \textit{activity index}. A larger $n$ indicates a steeper brightening rate (brightening/fading faster). The equation reduces to Equation~\ref{eq:h} when $n=2$. On average, comets have $n=4$, although this value can vary significantly among different comets and across various activity phases for the same comet.

For meteors, absolute magnitude is defined as the apparent magnitude of the meteor corrected to an altitude of 100~km at the observer's zenith.

\begin{BoxTypeA}[box:name]{}

\section*{Nomenclature of minor planets}
Minor planets with well-determined orbits are assigned by a permanent catalog number in chronological order, followed by a name if one has been given, such as (90377) Sedna. Newly-discovered minor planets with only preliminary orbits are first assigned a unique provisional designation, which consists of an alphanumeric combination representing the year, half-month, and sequential number of the discovery. For example, 2003 VB$_{12}$ [for (90377) Sedna] is the 302th discovery (represented by alphanumeric combo B$_{12}$) made in the first half of November (represented by letter V), 2003. The provisional designation remains in use for permanently numbered objects, particularly for those that have not been named. For instance, 90377, Sedna or 2003 VB$_{12}$ all refer to the same object.

\section*{Nomenclature of comets}
Once discovered (or recovered, in the case of returning periodic comets), comets receive a unique provisional designation based on the year, half-month, and a sequential number. The provisional designation is preceded by one of the following prefixes: P/ for periodic comets, C/ for long (well over 200~yr) or non-periodic comets, or I/ for interstellar comets. Occasionally, X/, D/ or A/ are used for comets with poor orbits, have disappeared, or are likely asteroidal. In addition, periodic comets that have been observed over multiple returns are given a permanent number followed by ``P'', such as 1P/Halley. Comets are typically named after their discoverer(s).

Sometimes, an asteroidal object is later found out to be a comet. Such an object keeps its asteroidal designation but receives a cometary prefix. For example, P/2019 LM$_4$ (Palomar) was initially discovered as asteroidal 2019 LM$_4$ and was found to be a comet later.

A small number of objects exhibit characteristics of both asteroids and comets (more in \S~\ref{sec:types-continuum}). Such ``dual-status'' objects are sometimes assigned both minor planet and comet designations. Centaur object 95P/2060 Chiron, for instance, has both a comet designation (95P) and a permanent minor planet number (2060). This naming system has not been consistently applied owing to the ambiguity in recognizing these objects.

\section*{Meteor shower names}
Because meteoroids from the same meteor shower share the same orbits, they will appear to an Earth-based observer to radiate away from a \textit{radiant} point in the sky. Meteor showers are named after the nearest constellation or a bright star closest to the radiant at the peak of the shower, with a suffix of ``id'' added to the name of the constellation (e.g. $\delta$-Aquariids). Occasionally, the month of the peak of the shower is added to the name to distinguish showers with the same radiants active in different times of the year (e.g. October Draconids). In a singular case, the Quadrantid meteor shower is named after an obsolete constellation. At one time, meteor showers were also known after their parent bodies (e.g. October Draconids were also known as the Giacobinids, after comet 21P/Giacobini-Zinner), a practice that is obsolete. The naming of meteor showers is overseen by the IAU Working Group of Meteor Shower Nomenclature.

\section*{Naming of meteorites}
Meteorites are typically named after the location where they are found (e.g. the largest known meteorite, the Hoba meteorite, is named after the Hoba West farm in Namibia). Naming is overseen by the Nomenclature Committee of the Meteoritical Society.

\end{BoxTypeA}

\subsection{Asteroid-Comet Continuum}
\label{sec:types-continuum}

As the names suggest, appearance is a simple way to distinguish between asteroids and comets: asteroids appear star-like, while comets are fuzzy and sometimes exhibit a tail. Asteroids and comets can also be classified based on their compositions or orbital dynamics. Compositionally, comets contain ice (or ``volatiles'', materials with low vaporization points, such as water ice), whereas asteroids do not. Dynamically, asteroids have near-circular orbits, while comets have very elongated and sometimes parabolic or hyperbolic orbits. The orbital dynamics of a small body can be evaluated using the Tisserand parameter with respect to Jupiter, $T_\mathrm{J}$:

\begin{equation}
    T_\mathrm{J} = \frac{a_\mathrm{J}}{a} + 2 \left[ (1-e^2) \frac{a}{a_\mathrm{J}} \right]^{1/2} \cos{i}
\end{equation}

\noindent where $a$, $e$ and $i$ are the semimajor axis, eccentricity and inclination of the object-in-question, and $a_\mathrm{J}=5.2$~au is the semimajor axis of Jupiter. $T_\mathrm{J}$ indicates whether an object crosses Jupiter's orbit and provides a measure of its speed relative to Jupiter at their closest approach: Jupiter itself has $T_\mathrm{J}=3$, objects with $T_\mathrm{J} \lesssim 3$ cross Jupiter's orbit, while those with $T_\mathrm{J} \gtrsim 3$ do not. A larger difference to $T_\mathrm{J}=3$ indicates higher encounter speed. Typical main-belt asteroids have $3 \leq T_\mathrm{J} \leq 4$, Jupiter-family comets (JFC; see \S~\ref{sec:population-comet}) have $2 \leq T_\mathrm{J} \leq 3$, and comets with longer periods have $T_\mathrm{J}<2$. We note that Tisserand's equation was derived under restricted three-body dynamics; in reality, researchers often use slightly higher $T_\mathrm{J}$ (3.05 to 3.10) to distinguish comets and asteroids \citep{hsieh2016potential}.

\begin{figure}[t]
\centering
\includegraphics[width=0.7\textwidth]{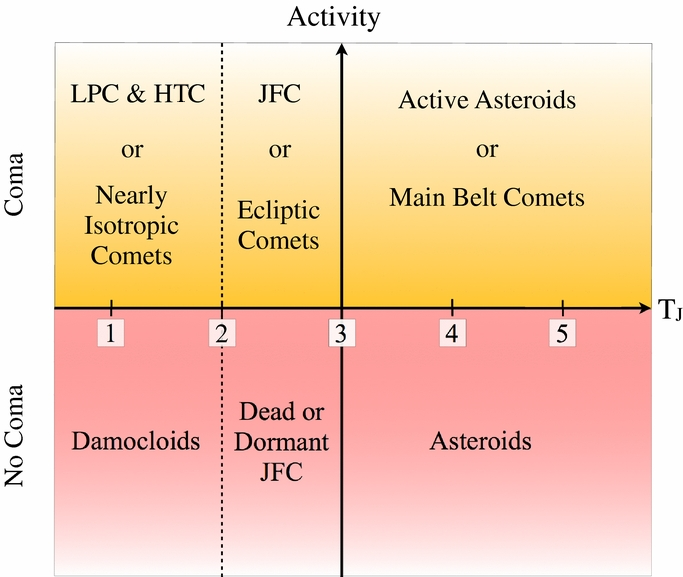}
\caption{A simple schematic diagram showing the classification of small bodies by the presence of the coma and $T_\mathrm{J}$. Acronyms are: LPC = long-period comet, HTC = Halley-type comets; JFC = Jupiter-family comets. Centaurs can appear anywhere on the diagram and are not shown. \S~\ref{sec:population} describes each population in greater details. From \citet{jewitt2012active}. \copyright AAS. Reproduced with permission.}
\label{fig:tj}
\end{figure}

These three approaches (based on appearance, compositional, or dynamical properties) usually yield consistent results: objects with a coma and ice signatures are usually on comet-like orbits, while objects on asteroid-like orbits do not exhibit a coma or ice signatures. However, a small number of objects discovered in recent years have challenged this paradigm. Dual-status object 133P/(7968) Elst-Pizarro, the first such outlier, was discovered as a main-belt asteroid but was found to exhibit a coma and a tail. Objects on asteroid-like orbits that exhibit comet-like activity are known as \textit{active asteroids}, while apparently inert objects on comet-like orbits are known as \textit{asteroids-on-cometary-orbits} (ACOs) and are sometimes presumed to be \textit{dormant comets}. Additionally, active asteroids in the main asteroid belt are sometimes known as main-belt comets; and ACOs on long-period orbits are known as the Damocloids. Figure~\ref{fig:tj} shows a simple classification diagram based on $T_\mathrm{J}$ and the presence of a coma. All these objects are collectively known as the continuum or transitional objects.

Objects can be driven into the continuum by a variety of processes, which can work alone or in concert with others \citep{jewitt2012active}. For instance, planetary dynamics can move objects closer to the Sun, activating shallow-buried ice and turning asteroid-looking objects into comets; depletion of near-surface ice can cause comets to become dormant; collisions between asteroids can produce comet-like debris tails; and the Yarkovsky–O'Keefe–Radzievskii–Paddack (YORP) effect can rotationally destabilize asteroids, leading to mass ejection. As of 2024, a few dozen active asteroids have been discovered. A couple thousand objects fall in the definition of ACO, although most are likely asteroidal interlopers rather than dormant comets.

\begin{BoxTypeA}[box:name]{}

\section*{The Yarkovsky and the Yarkovsky–O'Keefe–Radzievskii–Paddack (YORP) effects}

The Yarkovsky effect is the modification of the orbit of a small body caused by anisotropic thermal emission. This effect is most pronounced on objects ranging in size between 0.1~m and 10~km. The magnitude of Yarkovsky effect is very small: the Yarkovsky acceleration of near-Earth asteroid (99942) Apophis, for example, is $-2.9\times10^{-14}~\mathrm{au/day^2}$ -- 15 orders of magnitude smaller than Sun's gravity. But it acts steadily in one direction. Yarkovsky effect plays an important role in driving asteroids from the main asteroid belt to the near-Earth region \citep[][see also \S~\ref{sec:mb-orbits}]{bottke2006yarkovsky}.

Anisotropic thermal emission also creates a thermal torque that modify the spin state of the small bodies, known as the YORP effect. This effect can destabilize the object, leading to breakup in extreme cases. The YORP effect is the dominant mechanism that create binary (sometimes multiple) systems among small asteroids. The YORP effect on binary systems, or the BYORP effect, often couples with tides between the two components to produce complex and fast-evolving spin-orbit interactions.

\end{BoxTypeA}

\section{Populations}
\label{sec:population}

The Solar System contains two disk-like structures: the asteroid belt in the inner Solar System and the Kuiper belt just beyond Neptune's orbit. In addition, several sub-populations bridge these two structures as well as the innermost and outer regions of the Solar System, including Trojans, near-Earth objects (NEOs), Centaurs, and various comet populations. Finally, the interplanetary dust cloud is an end product of some of these populations.

\subsection{Asteroid Belt}
\label{sec:population-mb}

\subsubsection{Overview}

The asteroid belt, also known the ``main asteroid belt'' or just ``main belt'', is a torus-shaped region occupying the wide gap between the orbits of Mars and Jupiter. This region contains about a million asteroids that are 1~km or larger, with the dwarf planet (1) Ceres being the largest ($D\sim940$~km) and most massive, containing 39\% of the belt's total mass. The next few largest asteroids -- (4) Vesta ($D\sim530$~km), (2) Pallas ($D\sim510$~km), and (10) Hygiea ($D\sim430$~km) -- contain 11\%, 9\% and 4\% of the belt's total mass, respectively. The total mass of the asteroid belt is estimated to be $(1.8\pm0.2)\times10^{-9} M_\odot$, or $0.06\%$ of Earth's mass, and is dominated by the few largest objects \citep{krasinsky2002hidden}. The asteroid belt is the remnant of a circumstellar disk that was once $100\times$ more massive than what it is today.

As of 2024, optical imaging surveys have cataloged most main belt asteroids larger than a few km. Astrometry (measurement of positions and motion of celestial objects) and broadband photometry (measurement of their brightness) provide knowledge of the orbits and sizes of observed asteroids. Surface properties and composition of main belt asteroids are primarily studied through spectroscopy in the optical and near-infrared wavelengths, although available data remains limited. Only $\sim0.5\%$ of main belt asteroids, mostly the brightest ones, have been spectrally classified. Still, this limited dataset has enabled us to explore the compositional diversity through the main belt and impact a wide range of scientific topics. Recent discoveries of main belt comets have raised questions about the origin and transport of water within the Solar System.

\subsubsection{Orbits}
\label{sec:mb-orbits}

\paragraph{Kirkwood Gaps}

\begin{figure}[t]
\centering
\includegraphics[width=\textwidth]{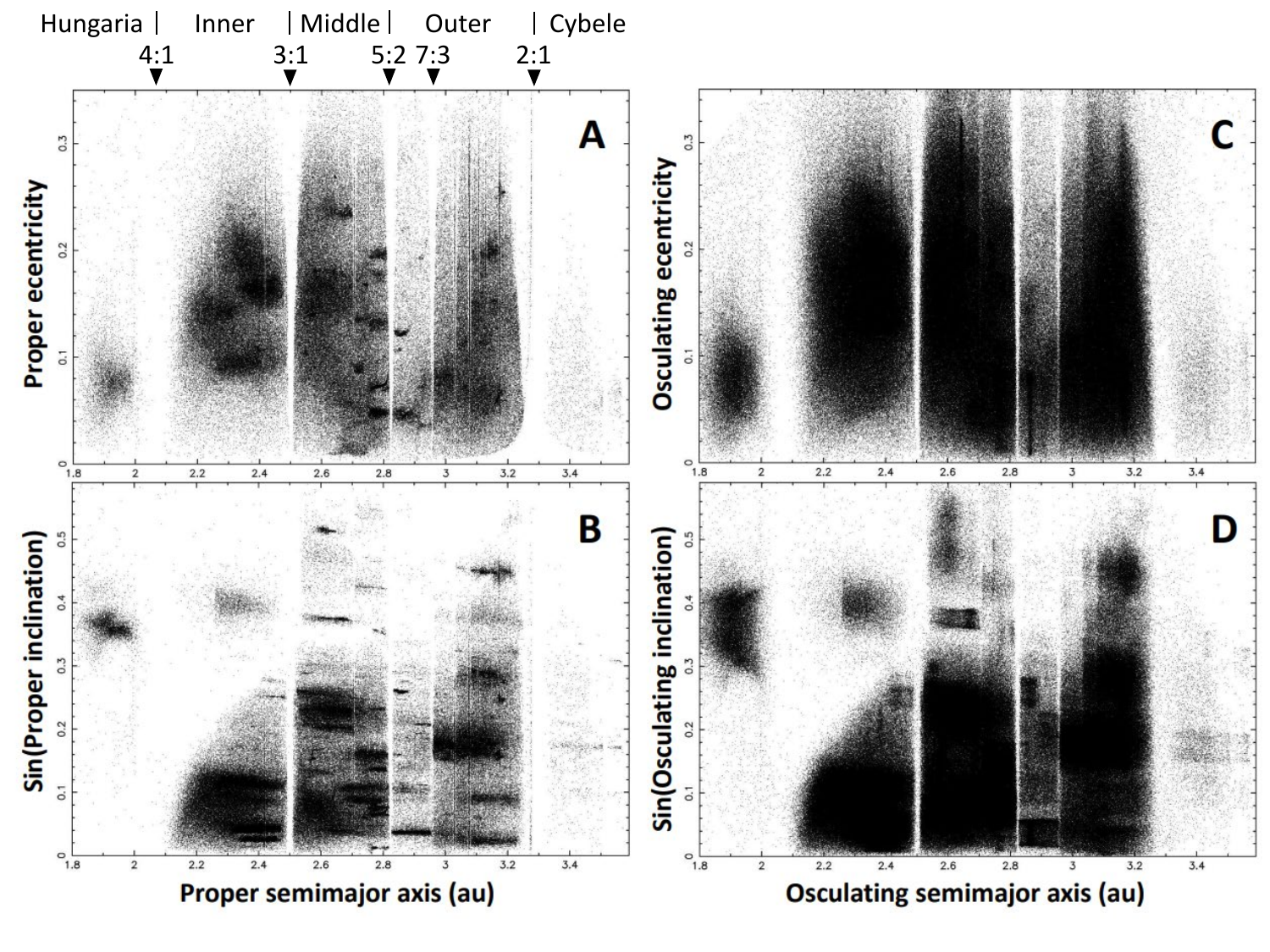}
\caption{Proper (panels A, B) and osculating (panels C, D) orbits of main belt asteroids. The concentrations in the proper orbits, which are mostly not visible in the osculating orbits, indicate the presence of asteroid families. Kirkwood gaps (mean motion resonances) and major regions of the main belt are marked. Adapted from \citet{nesvorny2024catalog}.}
\label{fig:mb-orbits}
\end{figure}

Most main belt asteroids have low orbital eccentricities ($e<0.4$) and semimajor axes between $\sim2.1$ and 3.3~au (Figure~\ref{fig:mb-orbits}). Instead of a broad, continuous distribution, main belt asteroids cluster in specific regions separated by the \textit{Kirkwood gaps}. These are narrow regions where orbital resonances with Jupiter quickly drive asteroids into high-eccentricity orbits, causing them to cross the orbits of terrestrial planets. Asteroids can drift into Kirkwood gaps through the Yarkovsky effect (see \S~\ref{sec:types-continuum}). A resonance region is denoted by a whole number ratio between the orbital periods of an asteroid in that region and Jupiter, optionally followed by ``J'' to indicate Jupiter. For instance, a particle at the 3:1J resonance orbits the Sun three times during one orbit of Jupiter.

Beyond the 2:1J resonance ($a=3.279$~au), orbital resonances with Jupiter have a different outcome: they protect asteroids from close encounters with Jupiter. Consequently, enhancements exist within these resonances, including the Hilda group at 3:2J, the Thule group at 4:3J, and the Cybele group at 7:4J.

\paragraph{Asteroid Families}

Analyzing the \textit{proper orbital elements} of main belt asteroids reveals clusters of asteroids with similar long-term motions (Figure~\ref{fig:mb-orbits}). These groupings, known as \textit{asteroid families}, likely originated from common parent bodies due to collisions. Proper orbital elements differ from the more commonly used \textit{osculating} orbital elements by representing the invariable part of the orbit over extended periods ($\gg10^6$~yr). This makes them useful for exploring the intrinsic dynamics of a targeted asteroid, independent of short-term planetary perturbations.

As of 2024, more than 100 asteroid families have been identified; about one-third of known main belt asteroids belong to a family. Large asteroid families can contain hundreds to thousands of asteroids. Families usually take their name from the lowest-numbered member; for example, the Vesta family derives its name from (4) Vesta.

The age of asteroid families can be estimated through the analyses of their Yarkovsky drift rates or the orbital structure of the family. The oldest asteroid families are $\sim1$~Gyr old, with the oldest being the Themis family (2-4 Gyr old). Families older than this have likely dissipated due to planetary dynamics and collisions. Conversely, the Karin and Veritas families are identified as the youngest major families, with ages of 5-8~Myr. Additionally, some smaller families with ages of $<1$~Myr have been proposed.

Members from the same family typically exhibit nearly identical spectral characteristics, reflecting their common origin from undifferentiated primordial planetesimals. This spectral feature can be used to identify interlopers -- objects that share the same orbital space with a family but are not otherwise related to it. (1) Ceres, for example, is a prominent interloper in the Gefion family.

\subsubsection{Composition}

\begin{figure}[t]
\centering
\includegraphics[width=\textwidth]{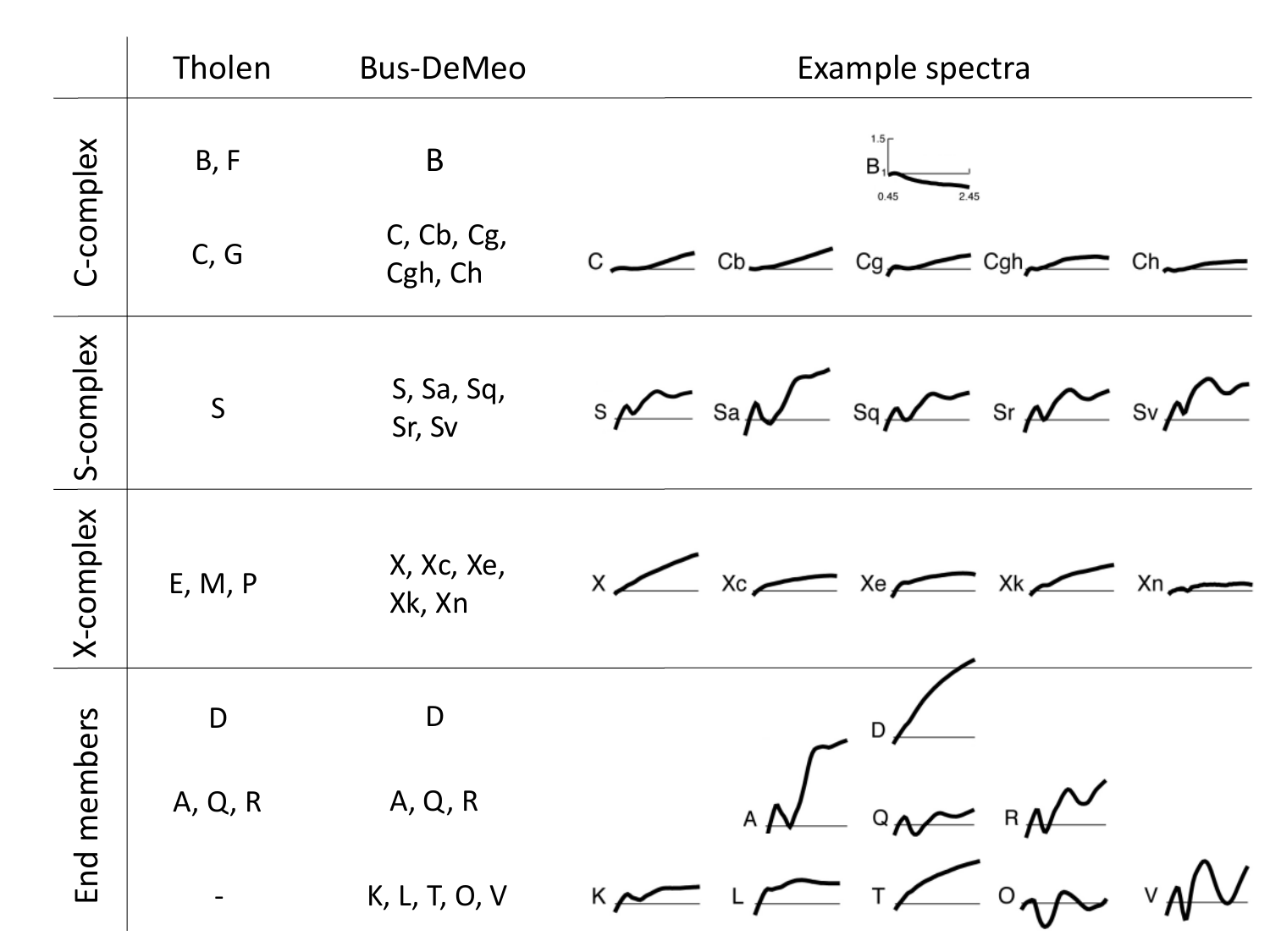}
\caption{Summary of the \citet{tholen1989asteroid} and Bus--DeMeo \citep{demeo2009extension} asteroid taxonomic systems. Compiled based on data from \citet[][Table~2]{cellino2002spectroscopic} as well as the Bus--DeMeo classification website (http://smass.mit.edu).}
\label{fig:asteroid-classes}
\end{figure}

Since asteroids do not emit visible light, reflectance spectroscopy is an important way for exploring their surface composition and properties. Since the 1970s, several taxonomic systems have been developed for optical spectrophotometry and spectroscopy (later extended into near-infrared) to classify asteroids, among which are the Tholen system \citep{tholen1989asteroid} and the Bus--DeMeo system \citep[][Figure~\ref{fig:asteroid-classes}]{demeo2009extension}. These systems differ in the details, but all contain two broad classes: ``C'' for carbonaceous objects, ``S'' for stony objects, plus a number of smaller populations that cannot be categorized as either C- or S-type. C-type asteroids exhibit generally featureless spectra, while S-type asteroids tend to display several absorption bands. A notable absorption feature near 1~\micron~is caused by silicates containing magnesium and iron. The asteroid taxonomic system roughly aligns with the meteorite classification system: C-type asteroids are thought to be the primary source bodies of carbonaceous chondrites, while S-type are associated with stony meteorites. Although carbonaceous chondrites represent only 5\% of known meteorites, C-type asteroids are the most common type of asteroids, accounting for $\sim75\%$ of all known asteroids. This disparity is attributed to terrestrial weathering of fallen meteorites \citep{bland2006weathering}.

\begin{figure}[t]
\centering
\includegraphics[width=\textwidth]{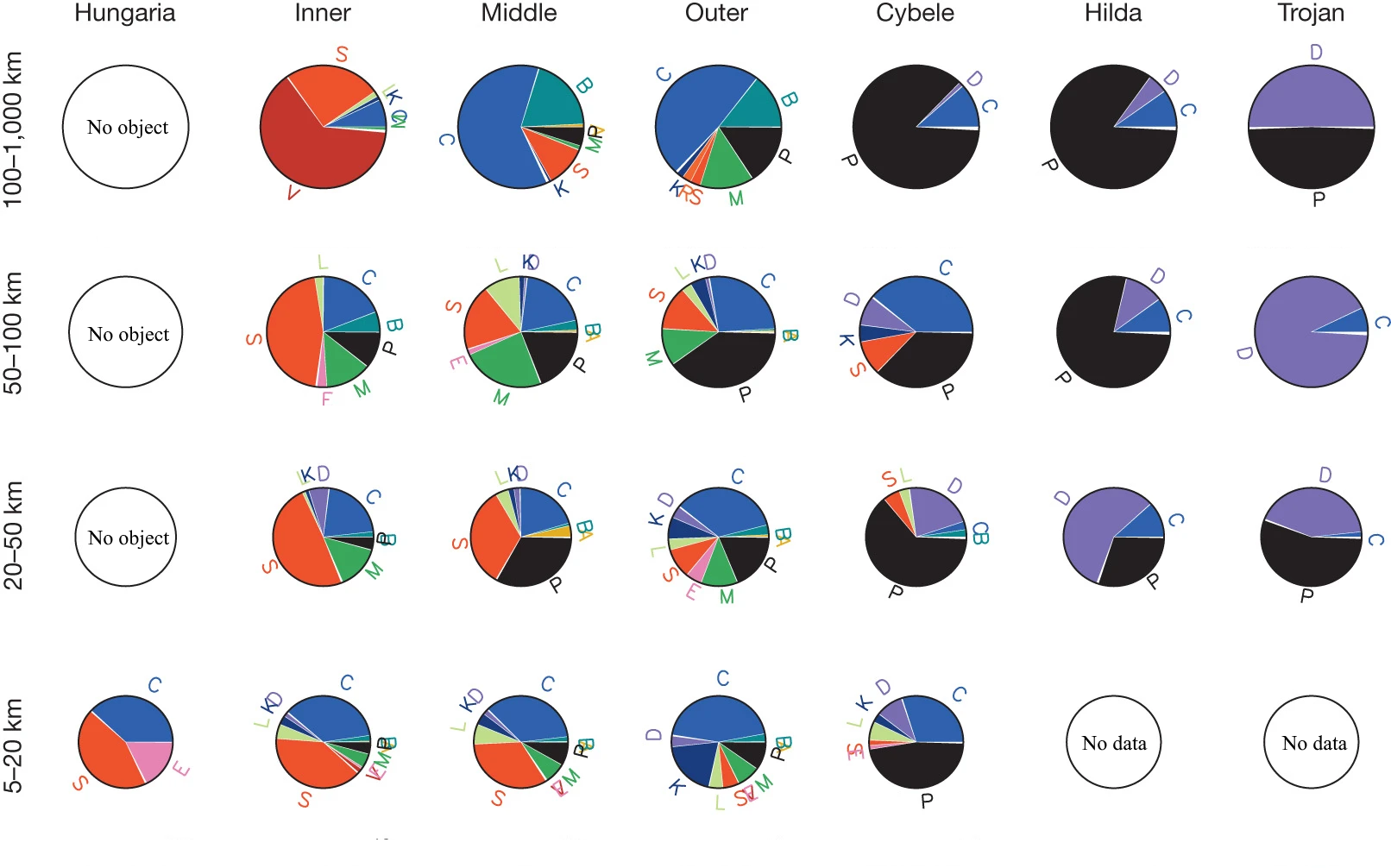}
\caption{Distribution of various asteroid spectral types across the asteroid belt and the Trojans, categorized by size. Adapted from \citet{demeo2014solar}.}
\label{fig:mb-comp}
\end{figure}

Systematic spectroscopic surveys have revealed a prominent compositional trend in the main belt (Figure~\ref{fig:mb-comp}): S-type asteroids dominate the inner main belt, while C-type and other primitive asteroids dominate the outer main belt. This trend is believed to reflect the conditions of the early Solar System, including the chemical and temperature gradient of the pre-solar nebula \citep{demeo2014solar}. The complexity of this picture suggests substantial modifications due to various evolutionary effects.

\paragraph{Water in the Asteroid Belt}

The discovery of water ice on (1) Ceres and several main belt comets implies that water may be common within the asteroid belt \citep{kelley2023spectroscopic}. While the asteroid belt is too warm for ice-rich objects to form in-situ, planetary migration models suggest that these objects may originate from the outer Solar System. Thermophysical models show that subsurface ices at asteroid belt distance can survive over the age of the Solar System \citep{schorghofer2008lifetime}. Buried ices in the polar regions of these objects can even make it to the near-Earth region assuming a stable spin orientation \citep{schorghofer2020preservation}, adding weight to the hypothesis that at least some of Earth's water comes from the asteroid belt.

\subsubsection{Formation and Evolution}

Like planets, the asteroid belt formed as part of the solar nebula. The widely accepted Nice model posits that giant planets originally formed much closer to the Sun, in which strong gravitational perturbation from a nearby Jupiter prevented planetary embryos in the region from growing into a full-fledged planet \citep{2001Icar..153..338P}. Despite this, one or multiple differentiated objects may have existed, as indicated by the metallic asteroids in the region. Subsequent planetary migrations dynamically excite the entire asteroid belt, removing $>99\%$ material within it within the first Myr of the formation of planets. Details such as the exact migration paths of Jupiter and Saturn are still being debated.

After the first Myr, the giant planets and the asteroid belt largely settled into their current positions. Larger asteroids occasionally collide, producing smaller fragments that contribute to the asteroid belt as well as the interplanetary dust cloud. Asteroids gradually drift across the belt due to the Yarkovsky effect until they reach resonance regions, where they are dynamically excited and may end up in planet-crossing orbits, contributing to the population of Near-Earth Objects (\S~\ref{sec:population-neo}).

\subsection{Jupiter Trojans}

The Jupiter Trojans, often referred to simply as Trojans, are a group of asteroids that share the same orbit with Jupiter. (The planets of Earth, Mars and Neptune also have their own Trojan population, but these are much smaller in size compared to Jupiter's Trojans.) Trojans are located at Jupiter's stable Lagrange points: the L$_4$ point, which is $60^\circ$ ahead of Jupiter in its orbit, and the L$_5$ point which is $60^\circ$ behind. They are relics of planetesimals that formed beyond Jupiter's orbit and were captured by Jupiter during planetary migration, at the same time the asteroid belt was formed \citep{morbidelli2005chaotic}.

Estimates suggest that the total mass of the Trojans is about 1/60th of the mass of the asteroid belt. A curious feature of the Trojans is that the number of L$_4$ Trojans is $\sim1.6\times$ more than that of L$_5$ Trojans. This asymmetry may be linked to Jupiter's outward migration during the early Solar System \citep{li2023asymmetry}.

Trojans have dark surfaces, with an average albedo of $\sim5\%$. Spectrally, they exhibit reddish, featureless spectra, with most members classified as D-type and a smaller number as P- or C-type asteroids, comparable to other outer main belt populations \citep{demeo2014solar} as well as Jupiter's irregular moons \citep{jewitt2004jupiter}. The Trojan population displays a spectral bimodality: a ``red'' cluster with a spectral slope of $\sim1\times10^{-4}~\r{A}^{-1}$, and a ``less-red'' cluster with $\sim6\times10^{-5}~\r{A}^{-1}$. This spectral behavior is commonly attributed to a combination of effects, including sublimation of ice, impact gardening, and space weathering \citep{emery2024surface}.

\subsection{Near-Earth Objects}
\label{sec:population-neo}

\subsubsection{Overview}

A \textit{Near-Earth Object} (NEO) is an asteroid or a comet with a perihelion of $<1.3$~au. Similarly, an asteroid with a NEO orbit is known as a near-Earth asteroid (NEA). Sometimes the term NEO is restricted to objects with $a<4.2$~au to exclude long-period comets. Not all NEOs come close to the Earth; for example, (529718) 2010 KY$_{127}$ has $q=0.29$~au but does not approach Earth closer than 0.7~au. A specific subset of NEOs, known as \textit{Potentially Hazardous Objects (Asteroids)} (PHOs/PHAs), pose a more direct threat to Earth. PHOs are defined as objects with a minimum orbit intersection distance (MOID) of $<0.05$~au and an absolute magnitude of $H<22$ (corresponding to a diameter of $D\gtrsim140$~m). These objects have the potential to cause significant damage to Earth in the event of an impact. None of the currently known PHOs are expected to collide with the Earth within the next century.

NEAs are further categorized into Amors, Apollos, Atens, and Atiras, based on their orbits relative to the Earth's:

\begin{itemize}
    \item Amors have orbits strictly outside Earth's orbit ($1.017$~au$<q<1.3$~au, where 1.017 au is Earth's aphelion distance);
    \item Apollos have $a>1$~au and Earth-crossing orbits ($q<1.017$~au);
    \item Atens have $a<1$~au and Earth-crossing orbits ($Q>0.983$~au, where 0.983~au is Earth's perihelion distance;
    \item Atiras have orbits strictly inside Earth's orbit ($Q<0.983$~au).
\end{itemize}

The Atira group also includes two sub-populations within the innermost region of the Solar System: the Vatiras (with orbits inside Venus's orbit) and the hypothetical Vulcanoids (with orbits inside Mercury's orbit). Asteroids can move between these groups after close encounters with Earth (and occasionally Venus). Atiras and Vatiras likely attained their current orbits due to inward gravitational scattering by Earth and Venus.

\begin{figure}[t]
\centering
\includegraphics[width=0.6\textwidth]{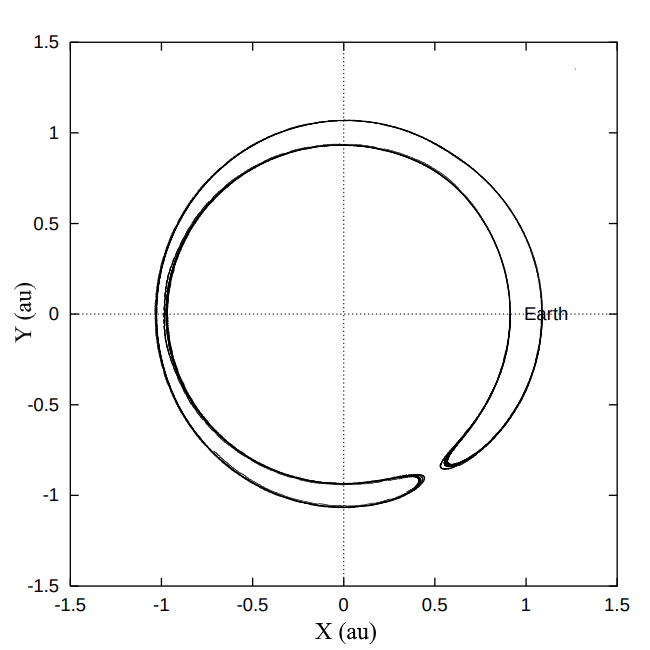}
\caption{Averaged mean motion of Earth co-orbiting asteroid (3753) Cruithne with respect to Earth in a corotating frame (such that Earth is fixed in the plot). Instead of librating around the fixed point like the Trojans, Cruithne follows a sophisticated horseshoe-like orbit. The deviation of Cruithne's radial distance from that of Earth is exaggerated by $30\times$ for clarity. From \citet{wiegert1998orbital}. \copyright AAS. Reproduced with permission.}
\label{fig:cruithne}
\end{figure}

Additionally, a small number of asteroids share Earth's orbit and are relatively dynamically stable. These include a few Earth Trojans, although some asteroids have complex horseshoe- or bean-shaped orbits around Earth that do not fit the definition of Trojans (e.g. Figure~\ref{fig:cruithne}). Venus also has a small population of co-orbital asteroids.

Comets constitute less than $1\%$ of the NEO population. Despite their small numbers, comets carry higher kinetic energy due to their highly elongated orbits and can cause greater damage in the event of an impact.

\subsubsection{Impact Risk}

\begin{figure}[t]
\centering
\includegraphics[width=\textwidth]{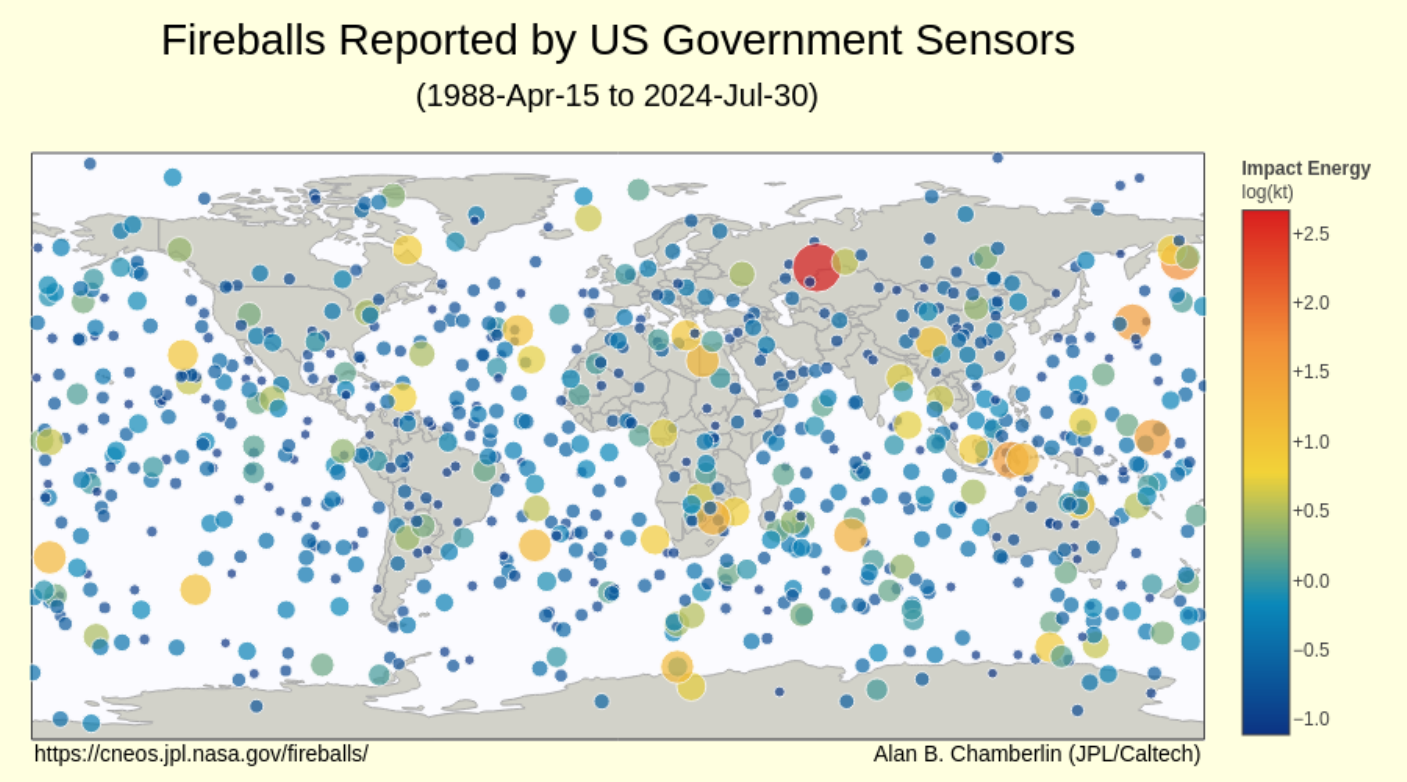}
\caption{Bolides and fireballs detected by geostationary satellites between 1988 and 2024. The largest data point in red is the 2013 Chelyabinsk event.}
\label{fig:jpl_fireball}
\end{figure}

Research into impact craters on Earth and other planetary bodies, as well as the impact of comet Shoemaker-Levy 9 (D/1993 F2) on Jupiter, has raised awareness of the hazards posed by NEOs. Since the 1990s, a number of wide-field telescopes have been dedicated to searching for and cataloging NEOs and other small Solar System bodies. The field of \textit{planetary defense} focuses on investigating and mitigating the risks posed by NEOs.

As of 2024, more than 30,000 NEOs have been discovered and cataloged. The first successful prediction of an asteroid impact occurred in 2008 with asteroid 2008 TC$_3$). Since then, another seven NEA impacts have been predicted before they occurred. All involving asteroids were a few meters in diameter, none of which reportedly cause damage on the ground. However, these represent only a small fraction of the asteroids that collide with Earth, primarily because nighttime telescopes have limited (though improving) sky coverage. Geostationary satellites, which have complete coverage of Earth's atmosphere, have recorded over 100 impact events since 1988, averaging about a few per year (Figure~\ref{fig:jpl_fireball}). The Chelyabinsk event in 2013, the largest impact event on Earth in over 100 years, was caused by a 18-m-sized asteroid approaching from the sunward direction. Meter-class impactors usually cause no significant damage. The Chelyabinsk event caused localized but significant damage, largely due to the asteroid's size and its proximity to a large city. Asteroids larger than $\sim20$~m can cause significant regional damage, and those larger than 100~m can cause global-level damage. Numerical models suggest that impacts of 100-m-class asteroids happen every $\sim5000$~yr on average, while collisions with km-class asteroids happen once every $\sim0.5$~Myr.

Efforts are ongoing to understand how to prepare for and potentially prevent destructive impacts. In 2022, the Double Asteroid Redirection Test (DART) mission successfully tested an asteroid deflection method by changing the orbit of Dimorphos, a satellite of NEA (65803) Didymos. Agencies and organizations such as the International Academy of Astronautics and the International Asteroid Warning Network are holding regular exercises and developing strategies with governments to prepare for potential asteroid impacts.

\subsubsection{Source Region and Dynamical Evolution}

\begin{figure}[t]
\centering
\includegraphics[width=0.8\textwidth]{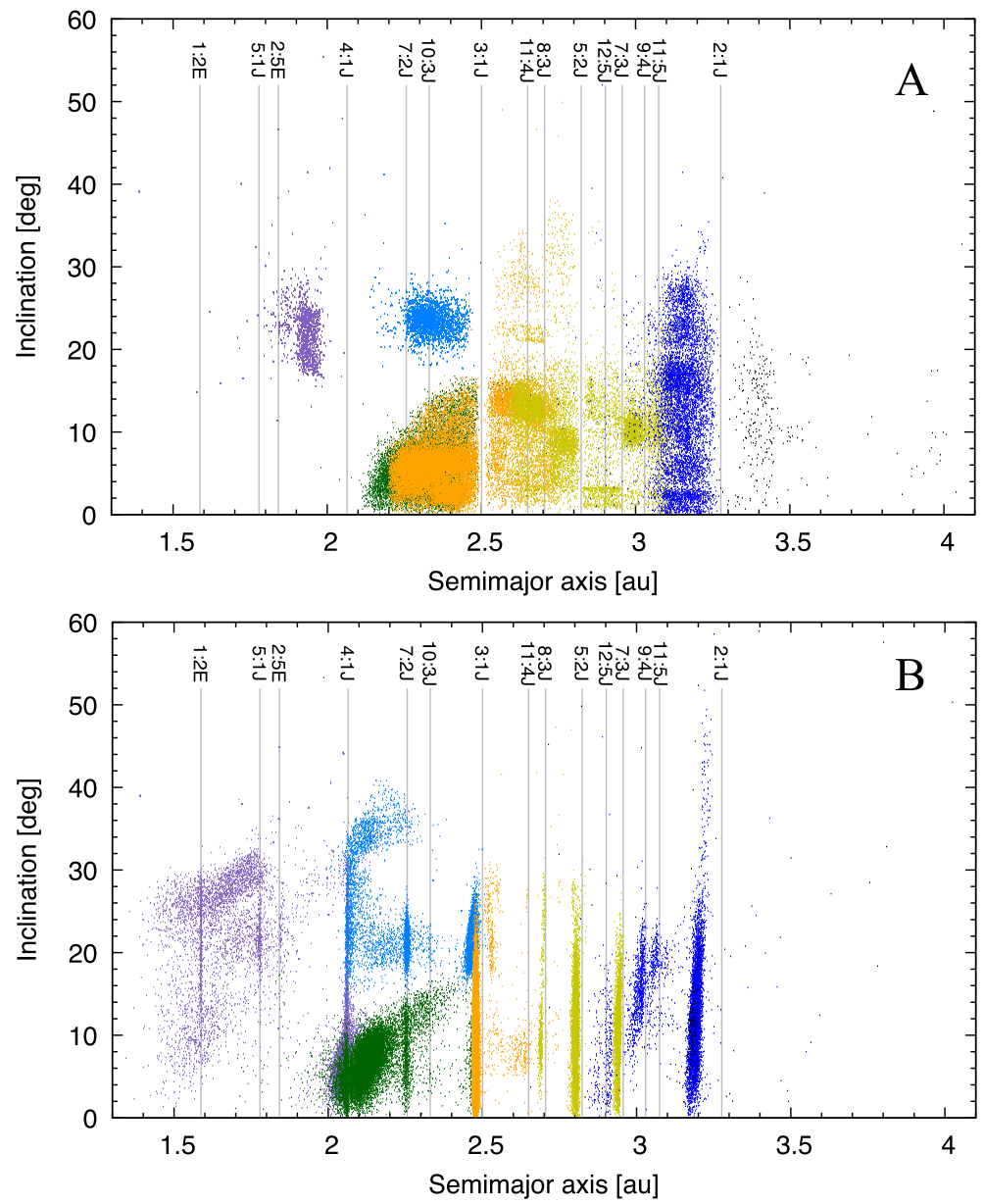}
\caption{Orbital distribution of test particles entering the NEO region: initial locations during a 100 Myr simulation (panel A) and orbital locations at the time of entry into the NEO region (panel B). Resonance zones are marked. Particles are color-coded based on their source regions. Reproduced from \citet{granvik2018debiased}.}
\label{fig:neo-source}
\end{figure}

As discussed in \S~\ref{sec:population-mb}, the NEO population is primarily supplied by castaway asteroids from the asteroid belt (Figure~\ref{fig:neo-source}). Dynamical models show that $\sim35\%$ NEOs origin from a region known as the $\nu_6$ resonance (a strong, inclination-dependent resonance zone in the inner asteroid belt that is gravitationally influenced by both Jupiter and Saturn), followed by the 3:1J resonance ($\sim25\%$), and the innermost region of the asteroid belt populated by Hungaria asteroids (another $\sim23\%$). The contribution of comets, primarily JFCs, accounts for only $\sim7\%$ of the NEO population \citep{granvik2018debiased}.

An object's source region also influences its dynamical evolution. Objects from the inner main belt are generally less susceptible to close encounters with Jupiter and are therefore more dynamically stable, with a higher likelihood of colliding with Earth or other terrestrial planets. While the average dynamical lifetime of NEOs is $\sim10^6$~yr, objects originating from the Hungaria region in the inner main belt have an average dynamical lifetime of $\sim37$~Myr, whereas objects from the 2:1J resonance in the outer main belt have $\sim0.4$~Myr; Hungaria-originated NEOs have a $\sim5\%$ chance to collide with the Earth while NEOs from the 2:1J resonance have a few $0.1\%$ \citep{granvik2018debiased}.

\subsection{Centaurs}

\subsubsection{Overview}
\label{sec:centaur-overview}

\begin{figure}[t]
\centering
\includegraphics[width=0.9\textwidth]{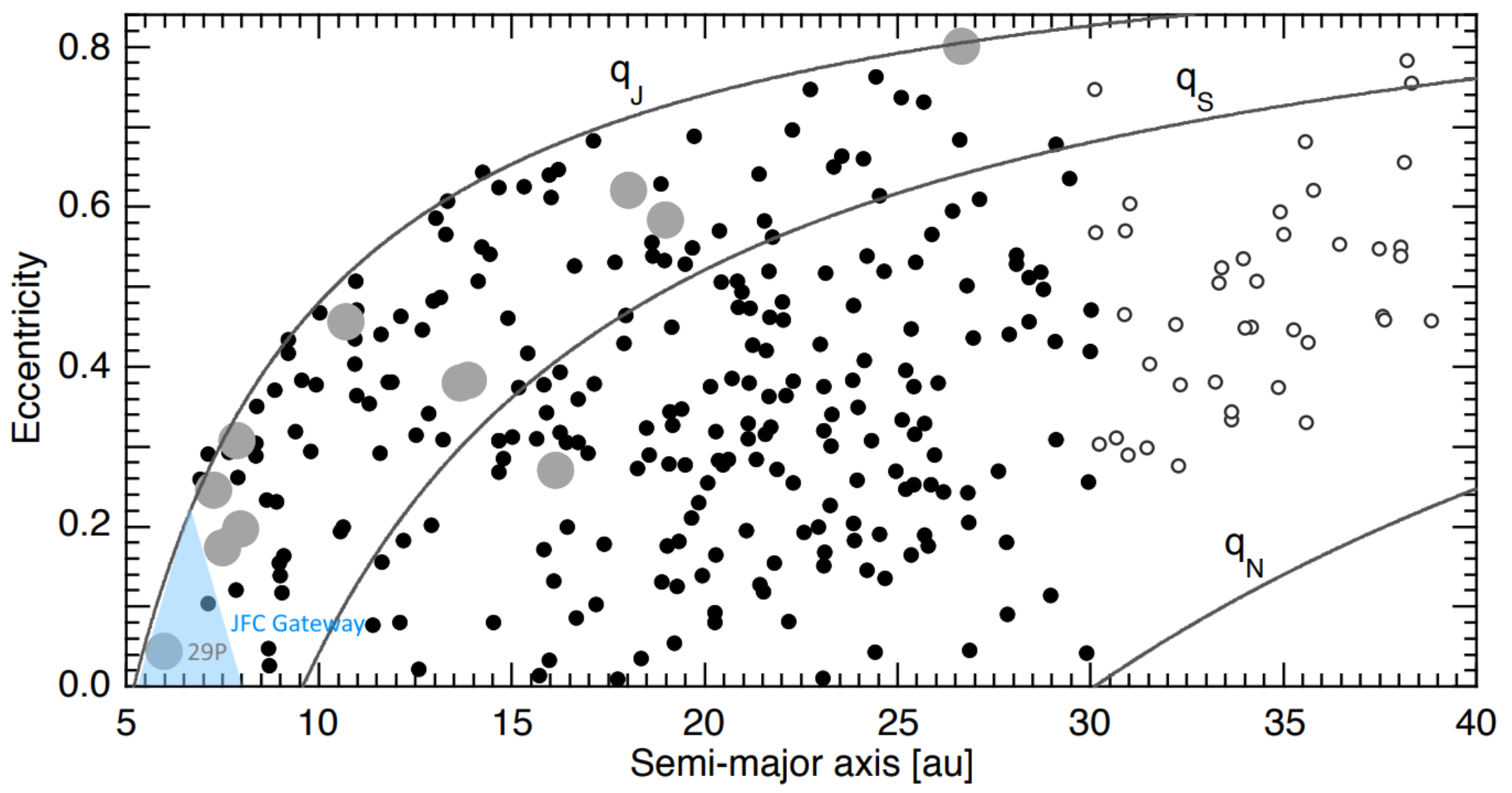}
\caption{The $a-e$ distribution of known Centaurs. Large gray dots are known active Centaurs, black dots are Centaurs under MPC definition, and hollow dots are objects exclusively classified by the Dark Energy Survey. The curves mark the orbits with perihelion equal to the semimajor axes of Jupiter ($q_\mathrm{J}$), Saturn ($q_\mathrm{S}$), and Neptune ($q_\mathrm{N}$). The blue shaded area marks the region of the JFC Gateway. Adapted from \citet{peixinho2020centaurs}.}
\label{fig:centaur_orb}
\end{figure}

Centaurs are small bodies that orbit the Sun between the orbits of Jupiter and Neptune (Figure~\ref{fig:centaur_orb}). They are ice-rich objects originated from the trans-Neptunian region perturbed into the planetary region. Recent estimates suggest there are $\sim10$ Centaurs with diameters of 100~km or larger, corresponding to a total population mass of $\sim1\%$ of the asteroid belt. Most Centaurs originate from the scattered disk in the trans-Neptunian region, where they can make close approaches to Neptune and be perturbed inward. These objects often cross the orbits of one or more of the giant planets, making them dynamically unstable. The dynamical lifetime of Centaurs is on the order of a few Myrs. Some Centaurs are further perturbed into the inner Solar System and become JFCs.

Centaurs were initially considered (inactive) minor planets similar to asteroids. The discovery of cometary activity on Centaur 95P/(2060) Chiron in 1989 changed this view. Survey data now show that about $10\%$ Centaurs display a coma at least occasionally. The name ``Centaur'' was derived from the first recognized Centaur, 95P/(2060) Chiron, named after the mythological centaur -- a creature with a half-human, half-horse body. This name also reflects the fact that astronomical Centaurs can exhibit characteristics of both asteroids and comets.

There are several definitions of Centaur. They generally agree on most objects, but differ on borderline objects:

\begin{itemize}
    \item The MPC defines Centaurs as any object with $q>5.2$~au and $a<30.1$~au, whereas 5.2~au and 30.1~au are the semimajor axes of Jupiter and Neptune.
    \item JPL defines Centaurs as objects with $5.5$~au$<a<30.1$~au that are not exclusively classified as comets, whereas 5.5~au is the aphelion of Jupiter. Apparent comets within the broadly-defined Centaur population are defined as Chiron-type comets ($a>5.2$~au and $T_\mathrm{J}>3$). This definition narrowly excludes 29P/Schwassmann-Wachmann~1 ($T_\mathrm{J}=2.99$) often regarded as a Centaur.
    \item The Deep Ecliptic Survey defines Centaurs as non-resonant objects whose osculating perihelia lie between the osculating semimajor axes of Jupiter and Neptune in the last 10~Myr \citep{elliot2005deep}.
\end{itemize}

As a result, certain objects may be classified as Centaurs under one definition but not in another. For example, (944) Hidalgo ($q=1.9$~au and $a=5.7$~au) is a Centaur under JPL classification but is typically considered a Damocloid.

Objects can transition between Centaur-like and JFC-like orbits multiple times during their lifetime in the planetary region. Comet 39P/Oterma, for instance, made a close approach to Jupiter in 1963, altering its orbit from $q=3.4$~au to 5.5~au, reclassifying it from a JFC to a Centaur.

Curiously, several large Centaurs have recently been found to possess ring systems, including (10199) Chariklo, 95P/(2060) Chiron, and possibly (54598) Bienor. Chariklo's ring system is relatively well-observed and contains at least two rings. Why and how rings can exist on these small, dynamically unstable bodies remains an open question.

\subsubsection{Physical Characteristics}

Centaurs are generally redder than most asteroids and comets found in the inner Solar System. Spectrophotometry (equivalent to ultra-low-resolution spectroscopy) has showed that Centaurs exhibit a bimodal color distribution, classified into two general categories: neutral/gray ($B-R\approx1.0$) and red ($B-R\approx2.0$). This bimodality is similar to that observed in the Trojan population and crudely correlates with the trans-Neptunian region from which the Centaurs originate. The color of a Centaur is linked to its dynamical history and activity: redder Centaurs tend to be younger and have smaller orbital inclinations, while nearly all active Centaurs are neutral in color.

Spectroscopic data is limited to the few brightest Centaurs. Like Trojans, Centaurs are spectrally largely featureless in optical. Important features exist in the near-infrared, where absorption features of H$_2$O can be found at 1.5, 1.65 and $2.0~\micron$, and methanol ice (CH$_3$OH) at $2.27~\micron$. Coma spectroscopy of a few active Centaurs has also revealed the presence of CO and CO$_2$ ices.

\paragraph{Comet 29P/Schwassmann-Wachmann~1 and the JFC Gateway}

Comet 29P/Schwassmann-Wachmann~1 is the brightest (though not the largest) object in the Centaur population. With an orbit just outside Jupiter's, 29P is likely on the verge of becoming a JFC, with a $65\%$ probability of doing so in the next $10^4$~yr \citep{sarid201929p}. It has a fairly large nucleus with a diameter of 60~km, and is characterized by frequent outbursts, averaging about once a month, which may be driven by cryovolcanism on its surface.

Dynamical models indicate that about a third of the Centaurs will eventually be perturbed into the inner Solar System and become JFCs \citep{tiscareno2003dynamics}. About two-thirds of these objects spend time in a low-eccentricity region just outside Jupiter's orbit, known as the ``JFC Gateway'' \citep[][see also Figure~\ref{fig:centaur_orb}]{sarid201929p}, although a significant fraction of these may be objects returning from JFC-like orbits \citep{guilbert2023gateway}. Objects in this region are dynamically highly unstable, only spending $10^3-10^4$~yr before being (re-)perturbed into JFC region. 29P is the most prominent Centaur in the Gateway region.

\subsection{Comets}
\label{sec:population-comet}

\subsubsection{Overview}

Comets are observationally classified as extended sources. A comet in its raw form consists only of a \textit{nucleus} composed of rock, ice, and dust, which is typically a few km in size. A fully developed comet includes several components: a slightly greenish \textit{coma} (due to emissions of molecular C$_2$ and C$_3$) surrounding the nucleus, a whitish \textit{dust tail} that can be curved and wide, and a straight, narrow, usually bluish \textit{ion tail} (primarily due to CO$^+$, though it may also contain neutral atoms). These structures vary in size: the coma is typically $10^4-10^5$~km in diameter, but can reach the size of the Sun for very large and/or active comets; ion tails of large comets can be several au long.

Comets are generally classified into two broad categories: short-period comets and long-period comets. These categories reflect different source regions and evolutionary histories, leading to some noticeable differences in their properties.

\subsubsection{Short-Period Comets: Jupiter-Family and Halley-Type}

\begin{figure}[t]
\centering
\includegraphics[width=0.6\textwidth]{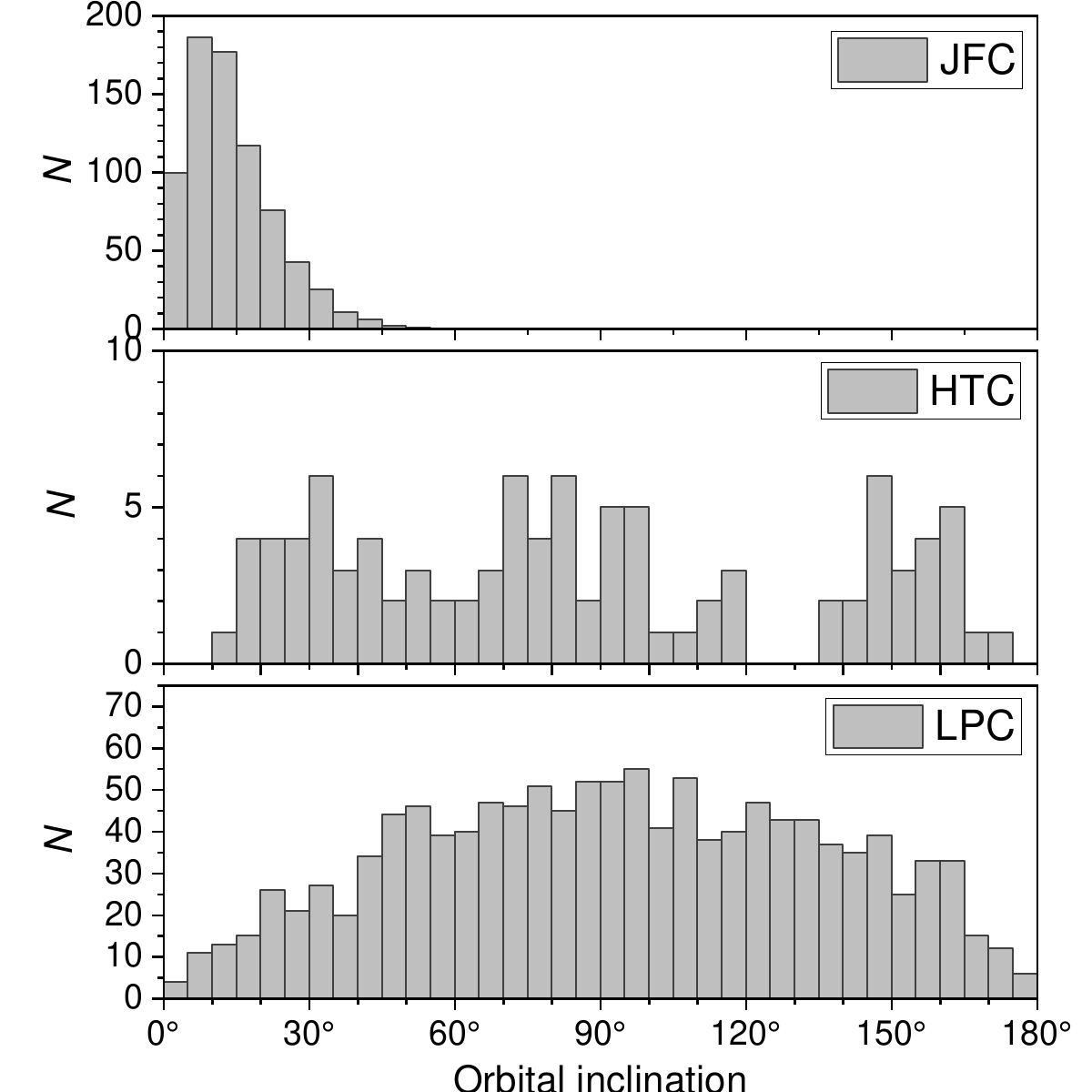}
\caption{Orbital inclination of different comet populations. JFCs are near the ecliptic plane, LPCs are nearly isotropic, and HTCs fall in between.}
\label{fig:comet-incl}
\end{figure}

Short-period comets (SPCs) have orbital periods $P<200$~yr. They are further divided into \textit{Jupiter-family comets} (JFCs) and \textit{Halley-type comets} (HTCs). In contrast to the concept of asteroid families that we discussed above, the word ``family'' in JFCs refers to their dynamical influence from Jupiter rather than physical relationship between the objects.

Traditionally, JFCs and HTCs are classified based on their orbital periods: JFCs have $P<20$~yr and HTCs have $20<P<200$~yr. However, simulations have shown that planetary dynamics frequently shuffle comets over the $P=20$~yr line \citep{levison1994long}. Thus, a $T_\mathrm{J}$-based scheme is often used: JFCs are comets with $2<T_\mathrm{J}<3$, while HTCs are comets with $T_\mathrm{J}<2$ while $P<200$~yr. Objects classified using the $T_\mathrm{J}$-based scheme tend to stay in the same classification throughout their lifetime. Additionally, \textit{ecliptic comets} -- those with orbits near the ecliptic plane (Figure~\ref{fig:comet-incl}) -- are defined as comets with $T_\mathrm{J}>2$, which crudely overlaps with the definition of JFCs.

Several sub-classes exist within SPCs. \textit{Encke-type comets} (ETCs), named after comet 2P/Encke, have $T_\mathrm{J}>3$ and $a<a_\mathrm{J}$. Although ETCs technically overlap with main-belt comets, objects like 2P/Encke are presumed to have different origins and evolutionary paths. Thus, ETCs typically exclude main belt comets. Chiron-type comets (discussed in \S~\ref{sec:centaur-overview}), with $T_\mathrm{J}>3$ and $a>a_\mathrm{J}$, are apparent comets among Centaurs.

JFC's low orbital inclination reminds us that they originate from a disk-like structure -- the scattered disk. JFCs frequently have close encounters with Jupiter and occasionally terrestrial planets, and are therefore dynamically unstable. The dynamical lifetime of JFCs is $\sim4.5\times10^5$~yr, during which their orbits are frequently modified. Most JFCs are eventually ejected from the Solar System, while a small but significant number ($\sim6\%$) impact the Sun. Their active lifetime is likely much shorter, as suggested by the small number of observed active JFCs and frequent disruption events among these comets. JFCs are likely only active for a few $10^3$~yr before disrupting or becoming dormant.

ETCs, a unique subset of broadly-defined JFCs, have aphelia well inside Jupiter's orbit, thus shielding them from close encounters with the planet. Given their presumed short active lifetime like other JFCs, this population needs to be frequently replenished by an unknown mechanism. Possible explanations include persistent non-gravitational acceleration or random coincidences.

The colors of JFC nuclei are similar to those of Centaurs \citep{jewitt2015color}, reflecting their close relationship. Active JFCs often appear bluer than inactive ones, possibly due to the blue scattering of cometary dust.

Chemically, many active JFCs display a depletion of carbon-chain molecules (such as C$_2$ and C$_3$) relative to other common cometary species like CN and OH \citep{a1995ensemble}. Observations of the fragmented JFC 73P/Schwassmann-Wachmann 3 revealed that this depletion extended throughout the interior of its progenitor \citep{russo2007compositional}. This depletion does not seem to correlate with the comet's dynamical age, suggesting it is likely a natal feature rather than one acquired through evolutionary processes.

As of 2024, about 800 JFCs and 50 ETCs have been cataloged. If we apply a criterion of $e>0.4$ to exclude main-belt comets, the number of ETCs drops to just 5. In addition, there are about 29,000 asteroids moving on JFC-like orbits, though most are likely asteroidal interlopers with $T_\mathrm{J}$ slightly below 3. A small subset of these asteroids may actually be dormant JFCs, though confirming their cometary nature is challenging. Some of these objects, such as (3552) Don Quixote, were recently found to exhibit extremely low levels of emissions, thereby revealing their cometary nature \citep{mommert2015exploreneos}.

HTCs exhibit a wider distribution of orbital inclinations compared to JFCs (Figure~\ref{fig:comet-incl}), suggesting that they did not primarily originate from the scattered disk. The exact source region of HTCs remains unclear, though recent dynamical models suggest that the edge of the scattered disk may be a likely region \citep{levison2006scattered}. These comets are thought to evolve into HTC-like orbits through the combined effects of galactic tides and planetary perturbations, potentially influenced by the Sun's orbit around the Milky Way Galaxy. Once on HTC orbits, these comets have a dynamical lifetime of $\sim1$~Myr, during which their orbits are heavily influenced by secular resonances with the giant planets.

As of 2024, about 100 HTCs have been cataloged, with most having only been observed during a single apparition. Due to their infrequent apparitions and lower numbers compared to JFCs and LPCs, HTCs have not been as extensively studied as the other two populations.

Damocloids, named after (5335) Damocles ($T_\mathrm{J}=1.15$), are asteroids on HTC-like orbits. Currently, about 180 Damocloids are known. These objects have colors similar to those of active Centaurs and JFC nuclei, and are generally believed to be dormant HTC nuclei \citep{jewitt2015color}.

\begin{BoxTypeA}[box:name]{}

\section*{Non-gravitational effect}

Comets release gas and dust into interplanetary space, which carries momentum and can subtly alter the orbit of the host object -- a phenomenon known as the ``non-gravitational effect''. This acceleration is very small, typically on the order of $10^{-10}~\mathrm{au/day^2}$ (or 11 orders of magnitude weaker than Sun's gravity). The magnitude and direction of this non-gravitational acceleration are highly variable over short timescales (such as a few orbits), usually rendering its long-term effects insignificant compared to the much larger gravitational perturbations from planets. However, in certain cases, persistent and intensive non-gravitational effects may be able to significantly influence a comet's long-term dynamical evolution. For instance, non-gravitation acceleration may play an important role in the formation of ETCs \citep{pittich2004jupiter}.

\end{BoxTypeA}

\subsubsection{Long-Period Comets (LPCs)}

LPCs are comets with $P>200$~yr. Comets on parabolic ($e=1$) or hyperbolic orbits ($e>1$) are sometimes referred to as non-periodic comets. As oppose to ecliptic comets, those with $T_\mathrm{J}<2$ are called \textit{nearly-isotropic comets} due to their nearly uniform distribution of orbital inclinations (Figure~\ref{fig:comet-incl}). This uniformity suggests that the structure where LPCs formed is spherical. Most hyperbolic comets, despite their hyperbolic orbits, actually originated within the Solar System. Their orbits become slightly hyperbolic due to planetary perturbations or non-gravitational accelerations, with only a few exceptions being truly ``interstellar'' comets that have $e\gg1$.

\begin{figure}[t]
\centering
\includegraphics[width=.8\textwidth]{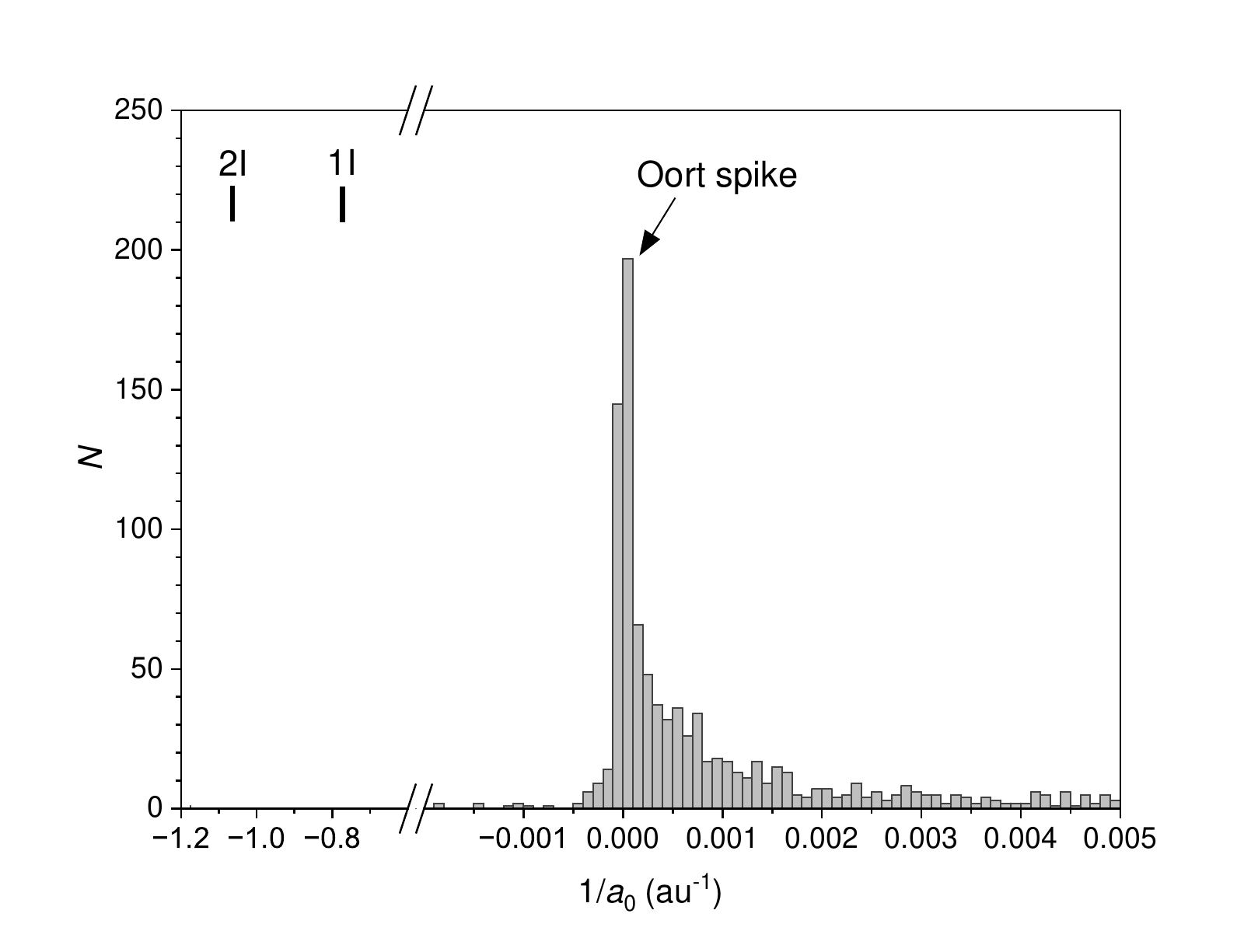}
\caption{Distribution of the inbound inverse semimajor axis $1/a_0$ of all LPCs as of 2024. The locations of the two known interstellar objects, 1I/`Oumuamua and 2I/Borisov, are also marked. Negative $1/a_0$ values correspond to hyperbolic orbits ($e>1$) and $1/a_0=0$ corresponds to parabolic orbits ($e=1$).}
\label{fig:lpc-a}
\end{figure}

Observed LPCs show an excess of comets with near-parabolic orbits, specifically those with an inbound inverse semimajor axis $1/a_0<10^{-4}~\mathrm{au^{-1}}$, known as the ``Oort spike'' (Figure~\ref{fig:lpc-a}). Based on this, \citet{1950BAN....11...91O} postulated the existence of a spherical \textit{Oort cloud}. (It is worth noting that the idea was earlier proposed by Ernst \"{O}pik in 1932 purely on dynamical grounds.) Comets with near-parabolic orbits that may be entering the planetary region for the first time are called \textit{dynamically new comets}. As they pass through the planetary region, these comets are perturbed by giant planets, scattering them into either lower-$Q$ orbits or hyperbolic orbits. A noted phenomenon in cometary science is the ``fading problem'', where there appears to be a depletion of returning comets with $1/a\gtrsim10^{-4}~\mathrm{au^{-1}}$. This suggests that comets must fade rather quickly within a few orbits in order to reproduce the observed $1/a$ distribution.

Oort cloud has not been directly observed; hence, most of what is known about it comes from studying LPCs. Planet migration models generally agreed that the Oort cloud objects were originally formed much closer to the Sun alongside planets, and were later scattered into the outer Solar System through gravitational interactions with giant planets \citep{vokrouhlicky2019origin}. Spectrophotometric observations indicate that both active and inactive LPCs exhibit colors similar to those of active JFCs and primitive asteroids, but distinct from inactive JFCs, Centaurs and other TNOs \citep{jewitt2015color}. This supports the idea that LPCs formed in the planetary region.

As of 2024, more than 600 LPCs (including non-periodic comets, excluding comet fragments) have been cataloged. Additionally, nearly 5,000 fragments of the Kreutz comet family (\S~\ref{sec:comet-sungrazer}) have been recorded. However, the studied LPCs are dominated by a handful of very bright comets, as most LPCs are faint. Almost all LPCs have only been observed during a single apparition, typically near their perihelia. Consequently, studies of LPC behavior often focus on population statistics rather than the long-term behavior of individual comets.

Long-term light-curves show that dynamically new comets tend to brighten slowly on their inbound legs. Some comets exhibit activity earlier than others, indicating the presence of ``supervolatiles'' such as CO ice \citep{meech2009activity}. Apart from a few outliers, the LPC population seems to be chemically homogeneous \citep{a1995ensemble}.

Asteroids on LPC-like orbits are referred to as \textit{Manx comets}, named after the Manx comet that has no tail. Identifying Manx comets is challenging as the coma (or the lack thereof) of distant objects is difficult to resolve. One such object, C/2014 S3 (PANSTARRS), displayed a spectrum compatible with rocky asteroids \citep{meech2016inner}. These objects are thought to be rocky material ejected into the Oort cloud by planetary migration in the early Solar System.

\subsubsection{Near-Sun Comets and Sungrazing Comets}
\label{sec:comet-sungrazer}

Comets with $q<0.3$~au are called \textit{near-Sun comets}. The traditional term of \textit{sungrazing comets} is now specifically defined as comets with $q<0.016$~au \citep[the fluid Roche limit of the Sun; cf.][]{jones2018science}. It is worth noting that the boundary between asteroids and comets becomes increasingly blurred at small solar distances, as intense solar heat can melt even rock and other refractory materials, causing otherwise inactive asteroids to exhibit comet-like behavior.

\begin{figure}[t]
\centering
\includegraphics[width=\textwidth]{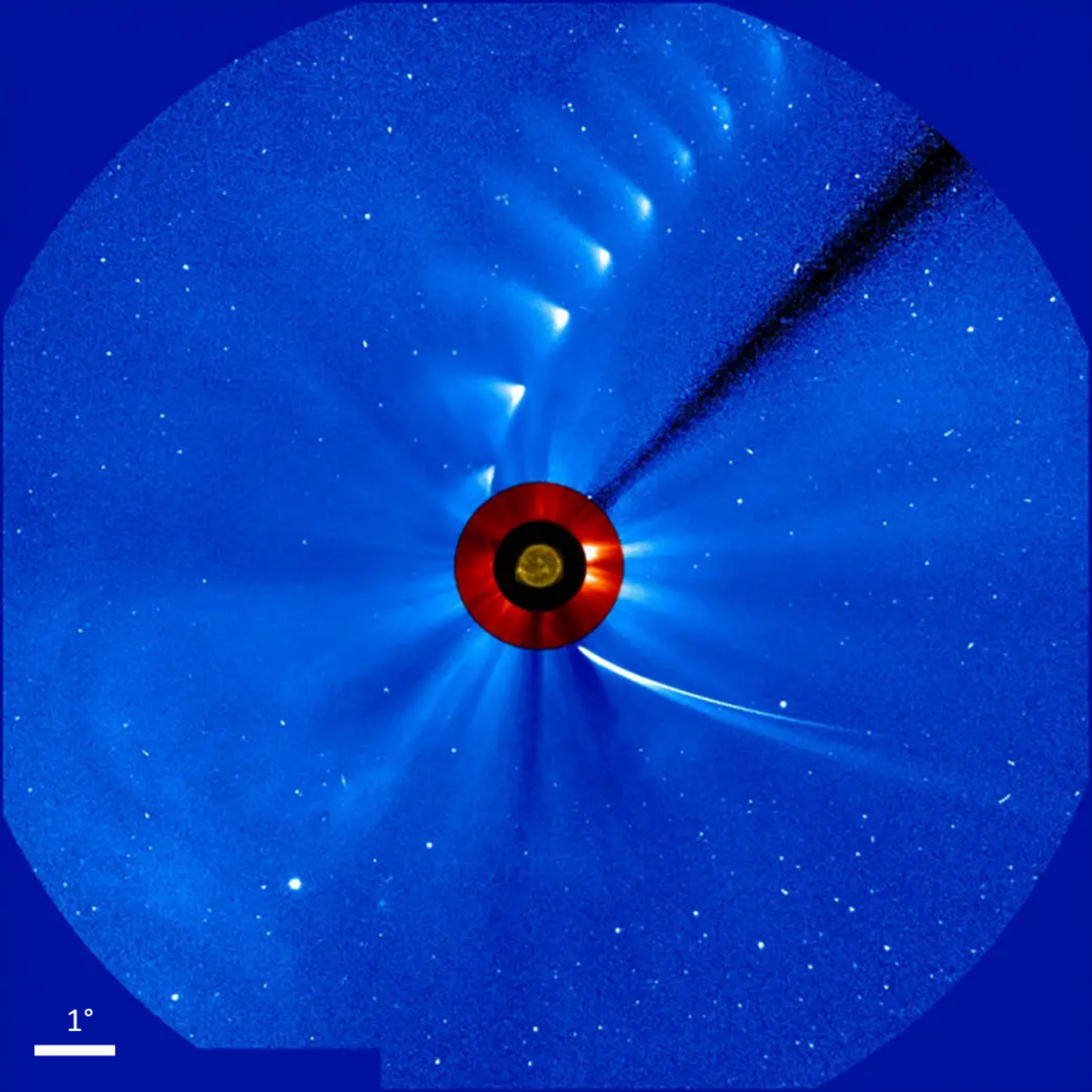}
\caption{Composite image of bright sungrazing comet C/2012 S1 (ISON) during its perihelion passage and subsequent disintegration in 2013. Individual images were taken by the Solar and Heliospheric Observatory (SOHO). The annuli from the Sun outwards are the field-of-views of SOHO's Extreme-ultraviolet Imaging Telescope (EIT), the Large Angle and Spectrometric Coronagraph (LASCO) C2 and C3 coronagraphs, respectively. Credits: ESA/NASA/SOHO/SDO/GSFC.}
\label{fig:ison}
\end{figure}

Fragmentation and disintegration are common among near-Sun comets (Figure~\ref{fig:ison}), often leading to the birth of new comet families. (Unlike JFCs, where the word ``family'' refers to dynamical similarities, the ``family'' here indicates physical relationships between members.) The Kreutz family of sungrazing comets is the most prominent comet populations in the near-Sun region, with perihelia almost at the surface of the Sun, or $q=0.0046$~au. These comets are descendants of a large, 50-km-class comet that disintegrated a few $10^3$~yr ago. Today, the system contains a few larger, km-class fragments and many thousands of deca-meter-class small fragments, most of which completely evaporate during perihelion passage. The Kreutz sungrazing comets are considered close analogs of the stargrazing comets observed in exoplanetary system. In addition to the Kreutz comets, several smaller families of near-Sun comets exist. However, it remains unclear whether some of those objects might actually be asteroidal in nature.

\subsubsection{Connection to Dust Trails and the Zodiacal Cloud}
\label{sec:comet-dust}

Comets deposit dust grains along their orbits, forming dust trails or, when emphasizing the link to meteor showers, meteoroid streams. These trails can be studied through their thermal emissions and meteor observations when Earth pass through them. Infrared imaging surveys show that dust trails are common among comets. Trails produced by JFCs blend into the broader interplanetary dust background over a few $10^3$~yr, though the timescale for individual objects can vary by a factor of several. Dust trails of HTCs and LPCs are less well-studied due to their smaller numbers, but meteor observations suggest that these comets can produce appreciable dust trails.

Infrared surveys and meteor data have also detected dust trails and meteoroid streams that are not associated with any known comets or active asteroids, often referred to as ``orphan trails''. These orphan trails are usually interpreted as remnants of disintegrated comets, though they could also belong to comets that have yet to be discovered, especially for objects on longer orbits. For example, the $\alpha$-Monocerotids meteor shower, which has an LPC-like orbit and occasionally produces strong meteor outbursts, has not been linked with any known comet.

\subsection{Interplanetary Dust Cloud}
\label{sec:zodi}

The interplanetary dust cloud, also known as the zodiacal cloud, is an extended, disk-like structure that spans the inner Solar System. It stretches from the innermost Solar System out to just beyond Jupiter's orbit. Asteroids and JFCs are the primary contributors to this dust cloud, though the relative contributions of each source remain unclear. Despite its total mass of the interplanetary dust cloud being relatively modest -- only $\sim10^{-6}$ of the asteroid belt's mass -- the spatial density of particles within the cloud is significant. At 1~au from the Sun, the volume density of particles of $\micron$-sized or larger is $\sim10^{-4}~\mathrm{m^{-3}}$. The interplanetary dust cloud is a significant source of foreground contamination across nearly all targets of observational astronomy, affecting wavelengths from optical wavelengths to far infrared.

The interplanetary dust cloud exhibits a bimodal size distribution: one peak in sub-mm sizes another in the \micron range. The sub-mm particles are primarily the result of larger asteroidal and cometary dust breaking down. These particles spiral inward towards the Sun due to the Poynting-Robertson drag. As they approach the Sun, they continue to break down and become part of the solar atmosphere known as the \textit{F-corona}. The remaining particles eventually reach a size where radiation pressure overcomes both Poynting-Robertson drag and solar gravity, sending these \micron-sized particles (known as $\beta$-meteoroids) onto hyperbolic orbits that escape the Solar System. This process creates the double-peak signature observed in the size distribution of the cloud, with exact peak locations varying slightly depending on the region within the cloud. Interstellar particles resembling $\beta$-meteoroids from other stars have also been observed in space.

In addition to the dust ``background'', the interplanetary dust cloud also includes dust trails or meteoroid streams from comets (\S~\ref{sec:comet-dust}), as well as particles from active asteroids and recent collisions that have created asteroid families (\S~\ref{sec:mb-orbits}). Planets can also capture and stabilize dust particles near their orbits, forming dust rings. Dust rings have been observed around the orbits of Mercury, Venus and Earth.

\section{Summary}
\label{sec:conclusion}

Small bodies can answer big questions. They offer profound insights on the formation and evolution of our Solar System, the origin of water and potentially life on Earth, and offer a unique perspective on similar structures observed in exoplanetary systems. Additionally, studying NEOs helps us understand potential threats to Earth and explore ways to mitigate possible impacts.

Originally rooted in astronomy, small body science has evolved into an interdisciplinary field, now spanning astronomy, atmospheric science, geology, and related disciplines in physics and chemistry. A wide range of techniques have been developed and employed to study small bodies, including Earth- and space-based telescopic observations, theoretical modeling, in-situ exploration with spacecraft, laboratory analysis of meteorites, returned samples and analog materials, as well as observations of meteors and related phenomena. Data obtained through these techniques enhance our understanding of small bodies from multiple perspectives.

The past few decades have been a golden era for small body science. Surveys have increased the small body catalog by nearly two orders of magnitude. Spectroscopic observations have created a broad but informative compositional map of the Solar System. Dynamical models have elucidated the evolution and interrelation of different classes of small bodies, offering insights to planetary migration in the early Solar System. Spacecrafts have explored small bodies ranging from near-Earth space to the Kuiper belt, with several missions returning samples that illuminate conditions in the early Solar System. The DART mission demonstrated asteroid deflection technology and created an artificially active asteroid. Advances in fireball surveillance and dynamical models have improved our ability to track meteorites to their source regions, deepening our understanding of Solar System's composition. Global meteor surveillance networks have helped map Earth's dust environment and improved our understanding of active bodies in near-Earth space. The discovery of two macroscopic interstellar objects has provided a glimpse into small bodies of extrasolar origin, hinting at the comparative hetereogeneity of planetary systems.

The future of small body science is promising. The Legacy Survey of Space and Time (LSST) is posed to increase the number of known small bodies by another order of magnitude. The NEO Surveyor mission will map the inner Solar System in the mid-infrared, providing crucial information on the physical properties of virtually all currently known small bodies. Other upcoming facilities, such as 30-m-class telescopes, the Roman Space Telescope, and SPHEREx mission, promise to advance small body science in their respective domains. The rich datasets generated by these facilities and missions will enhance theoretical models exploring the evolution of the Solar System. Of facilities in the smaller end, the widespread use of low-cost, off-the-shelf cameras and single-board computers will continue to expand global meteor surveillance, providing sensitive, uninterrupted data on the near-Earth dust environment and aiding in pinpointing the source regions of fallen meteorites. 

Several spacecraft missions have or will soon embarked to explore small bodies: Lucy will visit at least six Jupiter Trojans; OSIRIS-APEX and RAMSES will study PHA (99942) Apophis, which will make an extremely close (only 32,000~km above Earth's surface) but safe approach to Earth in 2029; the Psyche mission will visit metallic asteroid (16) Psyche, thought to be a protoplanetary core; Tianwen-2 will visit NEO (469219) Kamo`oalewa and main-belt comet 311P/PANSTARRS; DESTINY+ will visit active asteroid (3200) Phaethon; and the Comet Interceptor will investigate a yet-to-be-discovered LPC. Data from these new missions will provide clues to many outstanding questions in small body science and raise new ones.

\begin{ack}[Acknowledgments]
~This work is supported by NASA program 80NSSC22K0772. I thank Dimitri Veras and David Jewitt for helpful comments.
\end{ack}


\bibliographystyle{Harvard}
\bibliography{reference}

\end{document}